\documentclass[useAMS, usenatbib, a4paper]{mnras}
\usepackage{graphicx}
\usepackage{microtype}
\usepackage{xcolor}
\usepackage{fixltx2e}
\usepackage{booktabs}
\usepackage{siunitx}
\usepackage{color}
\usepackage{enumerate}
\usepackage{pdflscape}
\usepackage{rotating}
\usepackage{xr-hyper}
\usepackage{hyperref}

\usepackage[T1]{fontenc} 
\usepackage[utf8]{inputenc}

\usepackage{newtxtext}
\usepackage[varvw,smallerops]{newtxmath}

\usepackage{chemgreek}
\activatechemgreekmapping{newtx}
\usepackage{listings}

\hypersetup{colorlinks=True, linkcolor=blue!50!black, citecolor=black,
  urlcolor=blue!50!black}

\usepackage{etoolbox}
\robustify\bfseries
\robustify\itshape

\makeatletter
\patchcmd\@combinedblfloats{\box\@outputbox}{\unvbox\@outputbox}{}{%
  \errmessage{\noexpand\@combinedblfloats could not be patched}%
}%
\makeatother

\usepackage{bm}

\usepackage{aastex-compat}

\title
{Bow shocks, bow waves, and dust waves. III. Diagnostics}

\newcommand\AddressCRyA{Instituto de Radioastronom\'{\i}a y Astrof\'{\i}sica,
  Universidad Nacional Aut\'onoma de M\'exico, Apartado Postal 3-72,
  58090 Morelia, Michoac\'an, M\'exico}
\author[Henney \& Arthur]{
  William J. Henney\thanks{w.henney@irya.unam.mx}
  \& S. J. Arthur\\
  \AddressCRyA
}

\date{Accepted XXX. Received YYY; in original form ZZZ}

\pubyear{2019}

\newcommand\Qp{\ensuremath{Q_{\text{p}}}}

\newcommand{\alfven}{\ensuremath{_{\scriptscriptstyle\text{A}}}}

\newcommand{\wind}{\ensuremath{_{\text{w}}}}
\newcommand{\trap}{\ensuremath{_{\text{abs}}}}
\newcommand{\ke}{\ensuremath{_{\text{kin}}}}

\newcommand\rad{\ensuremath{_{\text{rad}}}}
\newcommand\Lya{\ensuremath{_{\text{Ly}\alpha}}}

\newcommand\sound{\ensuremath{c_{\text{s}}}}

\newcommand\mmp{\ensuremath{_{\text{\tiny MMP83}}}}
\newcommand\Hab{\ensuremath{_{\text{\tiny Habing}}}}
\newcommand\IR{\ensuremath{_{\text{IR}}}}
\newcommand{\thD}{\(\theta^1\)\,Ori~D}

\newcommand\shell{\ensuremath{_{\text{sh}}}}
\newcommand\M{\ensuremath{\mathcal{M}}}
\newcommand\hii{\ion{H}{ii}}

\newcommand\iftrap{\ensuremath{_{\text{trap}}}}
\newcommand\LOS{\ensuremath{_{\text{los}}}}

\begin{document}
\label{firstpage}
\pagerange{\pageref{firstpage}--\pageref{lastpage}}
\maketitle
\begin{abstract}
  Stellar bow shocks, bow waves, and dust waves all result from the
  action of a star's wind and radiation pressure on a stream of dusty
  plasma that flows past it.  The dust in these bows emits prominently
  at mid-infrared wavelengths in the range \SIrange{8}{60}{\um}. We
  propose a novel diagnostic method, the \(\tau\)--\(\eta\) diagram,
  for analyzing these bows, which is based on comparing the fractions
  of stellar radiative energy and stellar radiative momentum that is
  trapped by the bow shell.  This diagram allows the discrimination of
  wind-supported bow shocks, radiation-supported bow waves, and dust
  waves in which grains decouple from the gas.  For the wind-supported
  bow shocks, it allows the stellar wind mass-loss rate to be
  determined.  We critically compare our method with a previous method
  that has been proposed for determining wind mass-loss rates from bow
  shock observations. This comparison points to ways in which both
  methods can be improved and suggests a downward revision by a factor
  of two with respect to previously reported mass-loss rates.  From a
  sample of 23 mid-infrared bow-shaped sources, we identify at least 4
  strong candidates for radiation-supported bow waves, which need to
  be confirmed by more detailed studies, but no strong candidates for
  dust waves.
\end{abstract}


\section{Introduction}
\label{sec:introduction}

Bow-shaped circumstellar nebulae are observed around a wide variety of
stars \citep{Gull:1979a, Cox:2012a, Cordes:1993a} but are most
numerous around luminous OB stars \citep{Kobulnicky:2016a}.  They are
most commonly interpreted as bow shocks, due to a supersonic relative
motion of the surrounding medium, which interacts with the stellar
wind \citep{Wilkin:1996a}.  Early surveys for bow shocks around OB
stars \citep{van-Buren:1995a} concentrated on runaway stars
\citep{Blaauw:1961a, Hoogerwerf:2001a} that have been ejected at high
velocities (\(> \SI{30}{km.s^{-1}}\)) due to multi-body encounters in
star clusters or due to the core-collapse supernova explosion of a
binary companion.  However, only a small fraction of runaways show
detectable bow shocks \citep{Huthoff:2002a, Peri:2012a, Peri:2015a,
  Prisegen:2019a}. Targeted searches at mid-infrared wavelengths of
particular high-mass star forming regions such as M17 and RCW~49
\citep{Povich:2008a}, Cygnus~X \citep{Kobulnicky:2010a}, and Carina
\citep{Sexton:2015b} have revealed many bows around slower moving
stars.  In many cases, it may be streaming motions in the interstellar
medium that provide most of the relative velocity: weather vanes
rather than runaways \citep{Povich:2008a}.  More recently, large-scale
surveys of the Galactic plane \citep{Kobulnicky:2016a,
  Kobulnicky:2017a} and of nearby stars in the Bright Star Catalog
\citep{Bodensteiner:2018a} have revealed hundreds more such bow shocks.

Analytic and semi-analytic thin-shell models of bow shocks have been
developed \citep{van-Buren:1992a, Wilkin:1996a, Canto:1996},
including the effects of non-spherical winds and nonaxisymmetric bows
\citep{Wilkin:2000a, Henney:2002a, Canto:2005a, Tarango-Yong:2018a}.
Increasingly realistic numerical hydrodynamic simulations have been
performed \citep{Matsuda:1989a, Raga:1997a, Comeron:1998a, Arthur:2006a,
  Meyer:2014b, Mackey:2015a}, including magnetic fields
\citep{Meyer:2017a, Katushkina:2017a, Katushkina:2018a} and detailed
predictions of the dust emission \citep{Meyer:2016a, Acreman:2016a,
  Mackey:2016a}.


In \citet[Paper~I and Paper~II]{Henney:2019a, Henney:2019b} we
presented a taxonomy of stellar bows, which we divided into
wind-supported bow shocks (WBS) and various classes of
radiation-supported bows.  When the dust and gas remains well-coupled
(Paper~I), these are optically thin radiation-supported bow waves
(RBW) and optically thick radiation-supported bow shocks (RBS).  When
the dust decouples from the gas (Paper~II), inertia-confined dust
waves (IDW) and drag-confined dust waves (DDW) can result.

In Paper~I we derived expressions for the bow radius \(R_0\) in each of
the three well-coupled regimes as a function of the parameters of the
star and the ambient medium.  In Paper~II, we found the criteria for
the dust to decouple from the gas to form a separate dust wave outside
of the hydrodynamic bow shock.  The most important of these is that
the ratio of radiation pressure to gas pressure should exceed a
critical value, \(\Xi_\dag \sim 1000\).  

In order to provide an empirical anchor to our theoretical
calculations, we now consider how the parameters of our models might
be determined from observations.  The parameter-space diagrams, such
as Figure~2 of Paper~I, are not always useful in this regard, since in
many cases the ambient density and relative stellar velocity are not
directly measured.  Instead, we aim to construct diagnostics based on
the most common observations, which are of the infrared dust emission.
Key questions that we wish to address include
\begin{enumerate}[1.]
\item Can we distinguish observationally between radiation support
  (bow waves and dust waves) and wind support (bow shocks)?
\item Are there any clear examples of sources with radiation-supported bows?
\item In the case of wind support, can we reliably determine mass loss
  rates from mid-infrared observations?
\end{enumerate}

The outline of the paper is as follows. In
\S~\ref{sec:energy-trapp-vers} we describe how the optical depth and
the gas pressure in the bow shell can be estimated from a small number
of observed quantities.  In \S~\ref{sec:eta-tau-diagnostic} we place
observed bow sources on a diagnostic diagram of these two quantities
and discuss the influence of observational errors and systematic model
uncertainties. In \S~\ref{sec:cand-radi-supp} we use the diagram to
identify some candidates for radiation-supported bows. In
\S~\ref{sec:grain-temp-emiss} we calculate the grain emissivity as a
function of the stellar radiation field around OB stars.  In
\S~\ref{sec:stellar-wind-mass} we compare two different methods of
estimating the stellar wind mass loss rates from observations of the
bows. In \S~\ref{sec:discussion} we discuss various groups of bows
that require special treatment due to their diverse physical
conditions. In \S~\ref{sec:conclusions} we summarise out findings.





\section{Optical depth and pressure of the bow shell}
\label{sec:energy-trapp-vers}

\begin{figure*}
  \centering
  \includegraphics[width=0.8\linewidth]{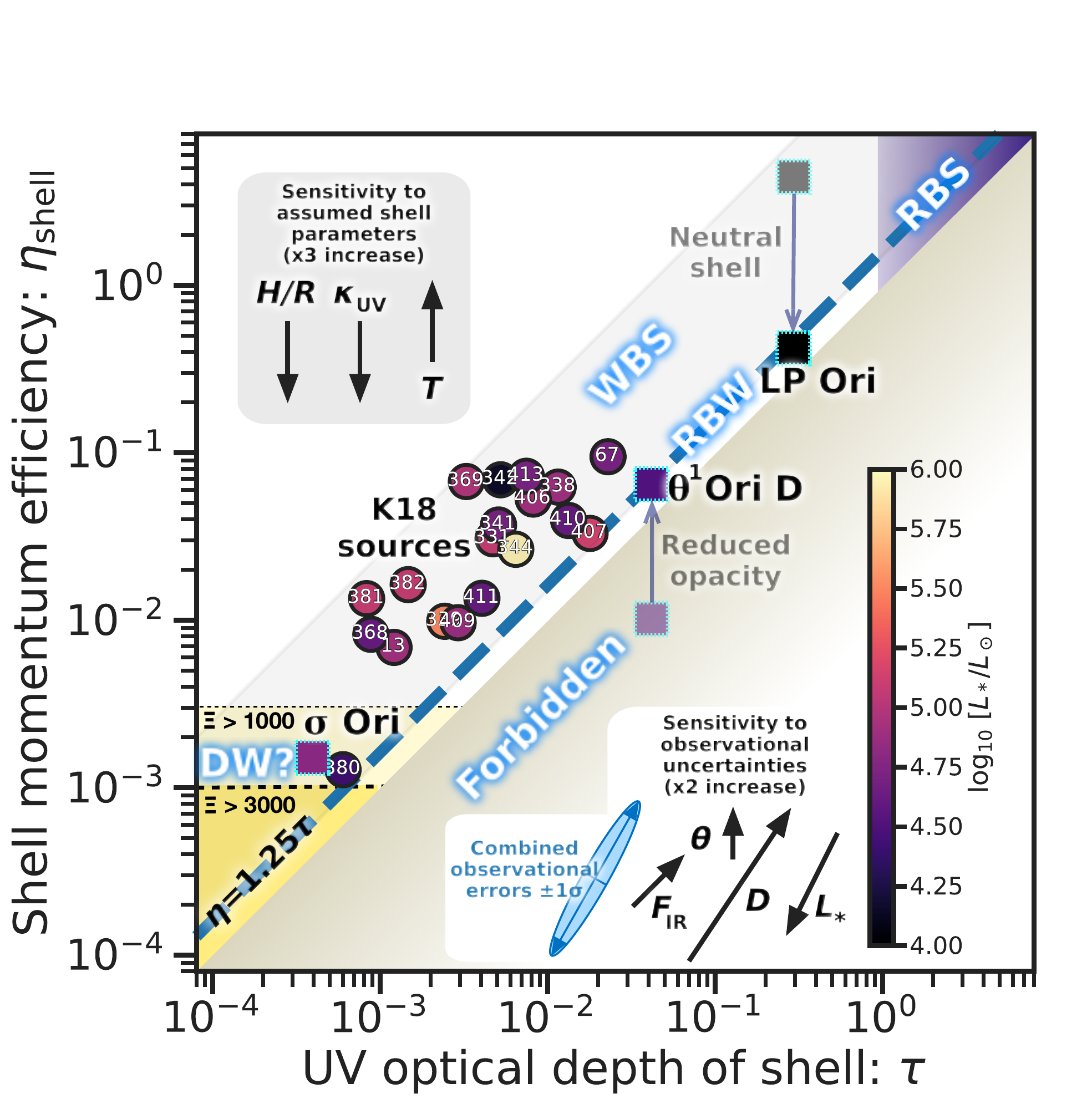}
  \caption[Observational diagnostic diagram]{Observational diagnostic
    diagram for bow shocks.  The shell optical depth \(\tau\) (\(x\)
    axis) and momentum efficiency \(\eta\shell\) (\(y\) axis) can be
    estimated from observations of the bolometric stellar luminosity,
    infrared shell luminosity, and shell radius, as described in the
    text.  Results are shown for the 20 sources (circle symbols) from
    \citet{Kobulnicky:2018a} plus three further sources (square
    symbols), where we have obtained the measurements ourselves (see
    Tab.~\ref{tab:observations}).  The color of each symbol indicates
    the stellar luminosity (dark to light) as indicated by the scale
    bar. The shell pressure is determined assuming a gas temperature
    \(T = \SI{e4}{K}\), an absorption opacity
    \(\kappa = \SI{600}{cm^2.g^{-1}}\), and a thickness-to-radius ratio
    \(H/R = 0.25\).  The sensitivity of the results to a
    factor-of-three change in each parameter is shown in the upper
    inset box.  Exceptions are the two Orion Nebula sources, \thD{}
    and LP~Ori, where the small dim squares show the results of
    assuming the standard shell parameters, while the large squares
    show the results of modifications according to the peculiar
    circumstances of each object, as described in the text.  The lower
    inset box shows the sensitivity of the results to a factor-of-two
    uncertainty in each observed quantity: distance to source \(D\);
    stellar luminosity \(L_*\), shell infrared flux \(F\IR\); shell
    angular size \(\theta\).  Lines and shading indicate different
    theoretical bow regimes (see Paper~I and Paper~II).  The dashed
    blue diagonal line corresponds to radiation-supported bows, while
    the upper left region corresponds to wind-supported bows.  The
    upper right corner (purple) corresponds to optically thick bow
    shocks, while the lower left corner (yellow) is the region where
    grain--gas separation \textit{may} occur, leading to a potential
    dust wave.  However, the existence of a dust wave in this region
    is not automatic, since it only includes one of the four necessary
    conditions (\S\S~4.4 and 5.1 of Paper~II). The lower-right region
    is strictly forbidden, except in case of violation of the
    assumption that dust heating be dominated by stellar radiation. }
  \label{fig:All-sources-eta-tau}
\end{figure*}

A fundamental parameter is the optical depth, \(\tau\), of the bow
shell to UV radiation, which determines what fraction of the stellar
photon momentum is available to support the shell (see
\S~2.1 of Paper~I).  But the same photons also heat the
dust grains in the bow, which re-radiate that energy predominantly at
mid-infrared wavelengths (roughly \SIrange{10}{100}{\um}) with
luminosity \(L\IR\).  Assuming that Ly\(\alpha\) and mechanical
heating of the dust shell is negligible (see
\S~\ref{sec:unimp-other-heat}) and that the emitting shell subtends a
solid angle \(\Omega\), as seen from the star, then the optical depth
can be estimated as
\begin{equation}
  \label{eq:tau-empirical}
  \tau = -\ln \left( 1 - \frac{4\pi}{\Omega} \frac{L\IR}{L_*} \right)
  \approx \frac{2 L\IR}{L_*} \ ,
\end{equation}
where the last approximate equality holds if \(\tau \ll 1\) and the shell
emission covers one hemisphere.\footnote{%
  Note that the \(\tau\) of Paper~I is not exactly the same as the
  \(\tau\) of equation~\eqref{eq:tau-empirical}, but is larger by a
  factor of
  \(Q_P / Q_{\text{abs}} = 1 + \varpi (1 - g)/(1 - \varpi)\), where
  \(\varpi\) is the grain albedo and \(g\) the scattering asymmetry.
  For standard ISM grain mixtures,
  \(Q_P / Q_{\text{abs}} = \text{\numrange{1.2}{1.3}}\) at EUV/FUV
  wavelengths.}

A second important parameter is the thermal plus magnetic pressure in
the shocked shell, which is doubly useful since in a steady state it
is equal to \emph{both} the internal supporting pressure (wind ram
pressure plus absorbed stellar radiation) \emph{and} the external
confining pressure (ram pressure of ambient stream).  The shell
pressure is not given directly by the observations, but can be
determined by the following three steps:
\begin{enumerate}[P1.]
\item \label{P1} The shell mass column (\si{g.cm^{-2}}) can be
  estimated from the optical depth by assuming an effective UV
  opacity: \(\Sigma\shell = \tau / \kappa\)
\item \label{P2} The shell density (\si{g.cm^{-3}}) can be found from
  the mass column if the shell thickness is known:
  \(\rho\shell = \Sigma\shell / h\shell\).  In the absence of other information, a
  fixed fraction of the shell radius can be used.  In particular, we
  normalize by a typical value of one~quarter the star--apex distance:
  \(h_{1/4} = h\shell / (0.25 R_0)\).  This corresponds to a Mach
  number \(\M_0 = \surd 3\) if the stream shock is radiative, or
  \(\M_0 \gg 1\) if non-radiative (see \S~3.2 of Paper~I). Further
  discussion is given in \S~\ref{app:bow-shock-data} below. 
\item \label{P3} Finally, the pressure (\si{dyne.cm^{-2}}) follows by
  assuming values for the sound speed and Alfvén speed:
  \(P\shell = \rho\shell (\sound^2 + \frac12 v\alfven^2) \).
\end{enumerate}
It is natural to normalize this pressure to the stellar radiation
pressure at the shell, so we define a shell momentum efficiency
\newcommand\pc{\ensuremath{_{\text{pc}}}}
\begin{equation}
  \label{eq:eta-shell}
  \eta\shell \equiv \frac{P\shell}{P\rad}
  = \frac{4\pi R_0^2\, (\sound^2 + \frac12 v\alfven^2)\, c\, \tau}{L_*\, \kappa\, h\shell}
  \approx 245 \frac{R\pc \, T_4 \, \tau}{L_4 \, \kappa_{600} \, h_{1/4}} \ , 
\end{equation}
where \(c\) is the speed of light. In the last step we have assumed
ionized gas at temperature \(\num{e4}\,T_4\,\si{K}\) with negligible
magnetic support (\(v\alfven \ll \sound\)) and written the stellar
luminosity and shell parameters in terms of typical values, which we
summarize below.  Note that the shell momentum efficiency is simply
the reciprocal of the radiation parameter from Paper~II's
equation~(23): \(\eta\shell = \Xi\shell^{-1}\), which provides yet a third
use for \(\eta\shell\), since \(\Xi\) is paramount in determining whether
the grains and gas remain well-coupled (see \S~4.4 of Paper~II).

In this section and the remainder of the paper, we employ
dimensionless versions of the stellar bolometric luminosity, \(L_*\),
wind mass-loss rate, \(\dot{M}\), and terminal velocity, \(V\wind\),
together with the ambient stream's mass density, \(\rho\), relative
velocity \(v_\infty\), and effective dust opacity, \(\kappa\).  These are
defined as follows:
\begin{align*}
  \dot{M}_{-7} &= \dot{M} / \bigl(\SI{e-7}{M_\odot.yr^{-1}}\bigr) \\
  V_3 &= V\wind / \bigl(\SI{1000}{km.s^{-1}}\bigr) \\
  L_4 &= L_* / \bigl(\SI{e4}{L_\odot}\bigr) \\
  v_{10} &= v_\infty / \bigl( \SI{10}{km.s^{-1}} \bigr) \\
  n &= (\rho / \bar{m}) / \bigl( \SI{1}{cm^{-3}} \bigr) \\
  \kappa_{600} &= \kappa / \bigl( \SI{600}{cm^2.g^{-1}} \bigr) \ ,
\end{align*}
where \(\bar{m}\) is the mean mass per hydrogen nucleon
(\(\bar{m} \approx 1.3 m_{\text{p}} \approx \SI{2.17e-24}{g}\) for solar
abundances).

\section[The eta-tau diagnostic diagram]
{\boldmath The \(\eta\shell\)--\(\tau\) diagnostic diagram}
\label{sec:eta-tau-diagnostic}

\begin{table}
  \centering
  \caption[Observational]{Key observational parameters for star/bow systems}
  \label{tab:observations}
  \begin{tabular}{l S S S}
    \toprule
    Star & {\(L_* / \SI{e4}{L_\odot}\)}
    & {\(L_{\text{IR}} / \si{L_\odot}\)} & {\(R_0 / \si{pc}\)} \\
    \midrule
    \thD & 2.95 & 620 & 0.003 \\
    LP~Ori & 0.16 & 240 & 0.01 \\
    \(\sigma\)~Ori & 6.0 & 15 & 0.12 \\[\smallskipamount]
    K18 Sources & \numrange{1.4}{87} & \numrange{8}{2800} & \numrange{0.02}{1.35} \\
    \bottomrule
  \end{tabular}
\end{table}

In Figure~\ref{fig:All-sources-eta-tau} we show the resultant
diagnostic diagram: \(\eta\shell\) versus \(\tau\).  The horizontal axis
shows the fraction of the stellar radiative \emph{energy} that is
reprocessed by the bow shell, while the vertical axis shows the
fraction of stellar radiative \emph{momentum} that is imparted to the
shell, either directly by absorption, or indirectly by the stellar
wind (which is itself radiatively driven).  Radiatively supported bows
(DW, RBW, or RBS cases) should lie on the diagonal line
\(\eta\shell = ( Q_P / Q_{\text{abs}}) \tau \approx 1.25 \tau\), where we have used
the ratio of grain radiation pressure efficiency to absorption
efficiency found in the FUV band for the dust mixture shown in
Paper~II's Figure~6.  Wind-supported bows should lie above this line
and no bows should lie below the \(\eta\shell = \tau\) line, since
\(Q_P\) cannot be smaller than \(Q_{\text{abs}}\).

We have calculated \(\eta\shell\) and \(\tau\) using the above-described
methods for the 20 mid-infrared sources studied by
\citet{Kobulnicky:2018a} (K18) and plotted them on our diagnostic
diagram.  Details of our treatment of this observational material are
provided in the following subsection.  In order to expand the range of
physical conditions, we have included three additional sources (data
in Table~\ref{tab:observations}): bows around \thD{}
\citep{Smith:2005a} and LP~Ori \citep{ODell:2001c} in the Orion
Nebula, which show larger optical depths, plus the inner bow around
\(\sigma\)~Ori, which illuminates the Horsehead Nebula and has previously
been claimed to be a dust wave \citep{Ochsendorf:2014b,
  Ochsendorf:2015a}.  Details of the observations of these additional
sources will be published elsewhere.

\subsection{Treatment of sources from \citeauthor{Kobulnicky:2018a}}
\label{sec:kobulnicky}

In a series of papers \citeauthor{Kobulnicky:2018a} provide an
extensive mid-infrared-selected sample of over 700 candidate stellar
bow shock nebulae (\citealp{Kobulnicky:2016a, Kobulnicky:2017a,
  Kobulnicky:2018a}, hereafter K16, K17, and K18).  For 20 of these
sources, reliable distances and spectral classifications are provided
in Table~5 of K17 and Tables~1 and~2 of K18. In this section, we
outline how we obtain \(\tau\) and \(\eta\shell\) from the data in
these catalogs, while further aspects of the
\citeauthor{Kobulnicky:2018a} material are discussed in
\S~\ref{app:bow-shock-data}.  As this paper was being written, we
became aware of a new mass-loss study that greatly expands on the
earlier results of K18 (H.~Kobulnicky, priv.~comm.).  The new study
includes 70 bow shock sources and incorporates Gaia distances and
proper motions, together with new optical/IR spectroscopy of the
central stars.  Apart from making use of an updated spectral
classification of one source (\S~\ref{sec:anomalous-b-star}), we have
elected to concentrate on only the published K18 sample in this paper,
but will address the expanded sample in future publications.

The UV optical depth of the bow shell is obtained
(eq.~[\ref{eq:tau-empirical}]) from the ratio of infrared shell
luminosity to stellar luminosity.  The inverse of this ratio is given
in Table~5 of K17, but we choose to re-derive the values since the
spectral classification of some of the sources was revised between K17
and K18.  Although K17 found the total shell fluxes from fitting dust
emission models to the observed SEDs, we adopt the simpler approach of
taking a weighted sum of the flux densities \(F_\nu\) (in \si{Jy}) in
three mid-infrared bands:
\begin{multline}
  \label{eq:total-ir-flux}
  F_{\text{IR}}  \approx \bigl[  2.4\,(F_8 \text{ or } F_{12})
    + 1.6\,(F_{22} \text{ or } F_{24})  \\
  + 0.51\,F_{70}\bigr]
  \,\times \SI{e-10}{erg.s^{-1}.cm^{-2}} \ ,
\end{multline}
where \(F_8\) is Spitzer IRAC \SI{8.0}{\um}, \(F_{24}\) is Spitzer
MIPS \SI{23.7}{\um}, \(F_{12}\) and \(F_{22}\) are WISE bands 3 and 4,
and \(F_{70}\) is Herschel PACS \SI{70}{\um}.  The weights are chosen
so that the integral \(\int_0^\infty F_\nu \,d\nu\) is approximated by the
quadrature sum \(\Sigma_k F_k\, \Delta\nu_k\), under the assumption that fluxes in
shorter (e.g., IRAC \SI{5.8}{\um}) and longer (e.g., PACS
\SI{150}{\um}) wavebands are negligible.  Shell fluxes are converted
to luminosities using the assumed distance to each source, and stellar
luminosities are taken directly from K18 Table~2, based on
spectroscopic classification and the calibrations of gravity and
effective temperature from \citet{Martins:2005a}.

\begin{figure}
  \centering
  \includegraphics[width=\linewidth]{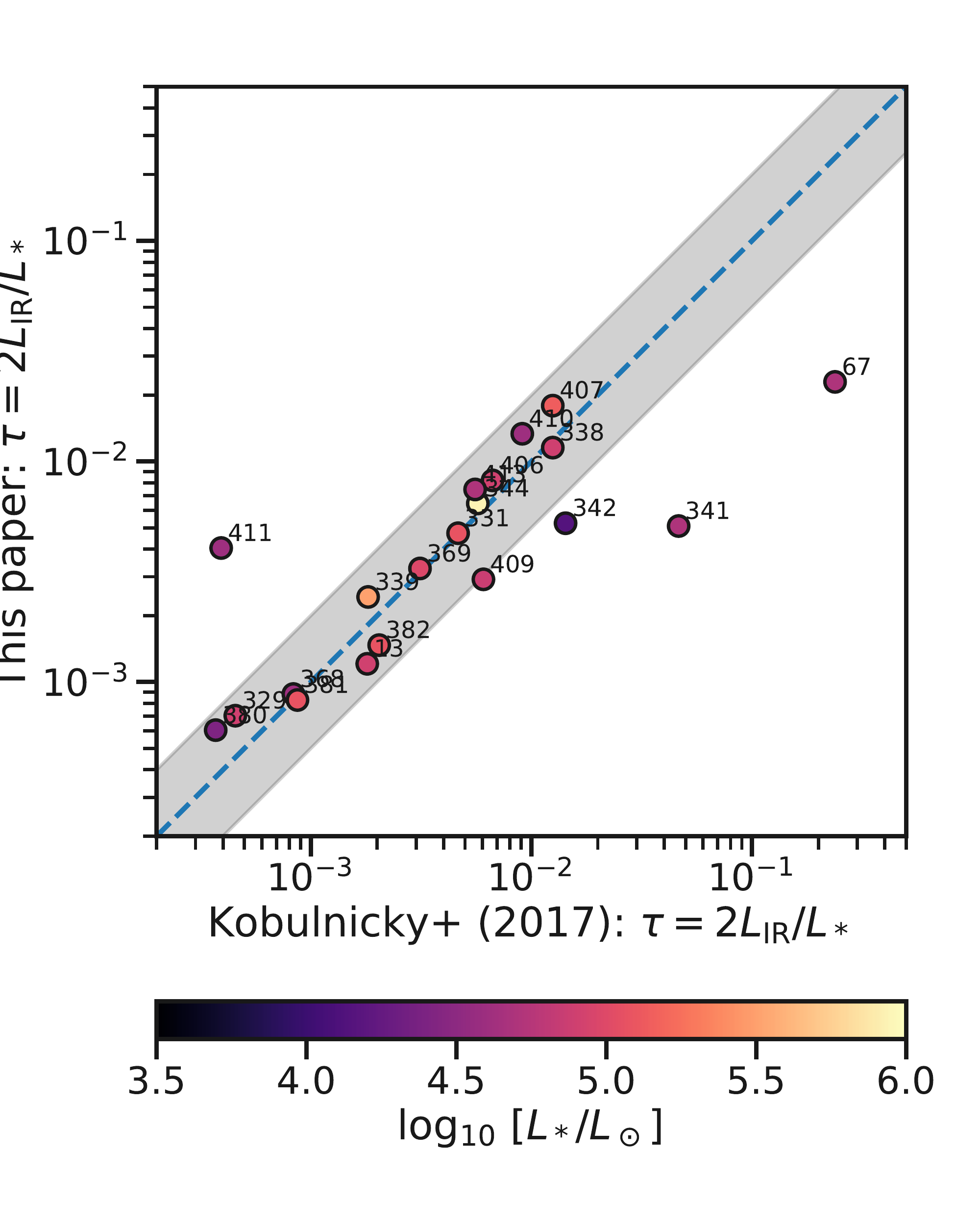}
  \caption{Comparison between shell-to-star luminosity ratios
    calculated as described in the text (\(y\) axis) with those given
    in K17 (\(x\) axis).  The blue dashed line signifies equality and
    the gray band shows ratios between 1/2 and 2.}
  \label{fig:k17-k18-comparison}
\end{figure}

In Figure~\ref{fig:k17-k18-comparison} we compare the \(\tau\) obtained
using the shell luminosity as described above with that obtained using
the luminosity ratios directly from K17 Table~5.  It can be seen that
for the majority of sources the two measurements are consistent within
a factor of two (gray band).  The four furthest-flung outliers can be
understood as follows:
\begin{description}
\item[\textit{Source 67}] This has a very poor-quality spectral fit in
  K17 (see lower left panel of their Fig.~12) and so 
  \(F_{\text{IR}}\) is overestimated by them by a factor of 10.
\item[\textit{Sources 341 and 342}] The spectral classes changed from
  B2V in K17 to O9V and B1V, respectively, in K18, increasing the
  derived \(L_*\), which lowers \(\tau\).
\item[\textit{Source 411}] The luminosity class changed from Ib (K17)
  to V (K18), so \(L_*\) has been greatly reduced, which increases
  \(\tau\).
\end{description}

\subsection{Random uncertainties due to observational errors}
\label{sec:rand-syst-uncert}

The fundamental observational quantities that go into determining
\(\tau\) and \(\eta\shell\) for each source are distance, \(D\); stellar
luminosity, \(L_*\); total infrared flux, \(F_{\text{IR}}\); and bow
angular apex distance, \(\theta\).  From these, the shell radius and
infrared luminosity are found as \(R_0 = \theta D\) and
\(L_{\text{IR}} = 4\pi D^2 F_{\text{IR}}\).  Rather than clutter the
diagram with error bars, we instead show the sensitivity to
observational errors in the lower-right box, where each arrow
corresponds to a factor of two increase (0.3~dex) in each quantity:
\(D\), \(L_*\), \(F_{\text{IR}}\), and \(\theta\).  We now calculate
uncertainty estimates for individual observational quantities that are
used in deriving not only \(\tau\) and \(\eta\shell\) but also mass-loss
rates, as discussed later in \S~\ref{sec:stellar-wind-mass}.

\subsubsection{Distance}
\label{sec:distance}

Most sources are members of known high-mass clusters with distance
uncertainties less than 20\% (0.08~dex). The only exception is
Source~329 in Cygnus, for which the distance uncertainty is roughly a
factor of~2 \citep{Kobulnicky:2018a}.

\subsubsection{Stellar luminosity}
\label{sec:stellar-luminosity}

The stellar luminosity is determined from spectral classification,
which makes it independent of distance.  Taking a \SI{2000}{K}
dispersion in the effective temperature scale \citep{Martins:2005a}
gives an uncertainty of 25\% in the luminosity, and adding in possible
errors in gravity and the effect of binaries, we estimate a total
uncertainty in \(L_*\) of 50\% (0.45~mag or 0.18~dex).

\subsubsection{Shell flux and surface brightness}
\label{sec:shell-flux-surface}

We estimate the uncertainty in shell bolometric flux,
\(F_{\text{IR}}\), by comparing two different methods: model fitting
\citep{Kobulnicky:2017a} and a weighted sum of the 8, 24, and
\SI{70}{\um} bands (eq.~[\ref{eq:total-ir-flux}]), giving a standard
deviation of 17\% (0.07~dex).  To this, we add the estimate of 25\%
for the effects of background subtraction uncertainties on individual
photometric measurements \citep{Kobulnicky:2017a}. The absolute flux
calibration uncertainty for both Herschel PACS \citep{Balog:2014a} and
Spitzer MIPS \citep{Engelbracht:2007a} is less than 5\%, which is
small in comparison. Combining the 3 contributions in quadrature gives
a total uncertainty of 0.12~dex.  We adopt the same uncertainty for
the \SI{70}{\um} surface brightness.

\subsubsection{Angular sizes}
\label{sec:angular-sizes}

For the angular apex distance, \(\theta\), the largest uncertainty for
well-resolved sources is due to the unknown inclination.
\citet{Tarango-Yong:2018a} show that the dispersion in true to
projected distances can introduce an uncertainty of \(30\%\)
(0.11~dex) in unfavorable cases (e.g., their Fig.~26).  For 5 of the
20 sources from \citet{Kobulnicky:2018a}, \(\theta\) is of order the
Spitzer PSF width at \SI{24}{\um}, so the errors may be larger.

\subsubsection{Stellar wind velocity}
\label{sec:stell-wind-veloc}

Although this is not strictly an observed quantity for the K18 sample,
we will treat it as such since it is estimated per star, based on the
spectral type.  K18 estimate 50\% uncertainty, and we adopt the same
here (0.18~dex).

\subsubsection{Combined effect of uncertainties on the
  \(\tau\)--\(\eta\shell\) diagram}
\label{sec:comb-effect-uncert}

Assuming that the uncertainty in each observational quantity is
independent, we can now combine them using the techniques described in
Appendix~\ref{sec:comb-uncert-covar} to find the \(\pm 1~\sigma\) error
ellipse, shown in blue in the figure.  It can be seen that
observational uncertainties in \(\tau\) and \(\eta\shell\) are highly
correlated: the dispersion is \SI{0.7}{dex} in the product
\(\eta\shell \tau\) but only \SI{0.16}{dex} in the ratio
\(\eta\shell/\tau\), with stellar luminosity errors dominating in both
cases.  Observational uncertainties are therefore relatively
unimportant in determining whether a given source is wind-driven or
radiation-driven, which depends only on \(\eta\shell/\tau\).  On the other
hand, they do significantly affect the question of whether a source
has a sufficiently high radiation parameter \(\Xi\) to possibly be a
dust wave.

\subsection{Systematic uncertainties due to assumed shell parameters}
\label{sec:syst-uncert-due}

A further source of uncertainty arises from the parameters of the
shocked shell that are assumed in steps P\ref{P1}--P\ref{P3}. Namely,
the relative shell thickness, \(h\shell/R_0\), the ultraviolet grain
opacity per mass of gas, \(\kappa\), and the shell temperature, \(T\).
These parameters effect only \(\eta\shell\), not \(\tau\), with a
sensitivity shown by arrows in the upper left box of
Figure~\ref{fig:All-sources-eta-tau}.

\subsubsection{Shell thickness}
\label{sec:shell-thickness}
For maximally efficient post-shock radiative cooling, the shell
thickness depends on the Mach number of the shock as
\(h / R_0 \sim M_0^{-2}\).  However, for ambient densities less than
about \SI{10}{cm^{-3}}, the minimum thickness is only about ten times
smaller than the \(h / R_0 = 0.25\) that we are assuming.  In
photoionized gas, this occurs at \(v \approx \SI{60}{km.s^{-1}}\),
corresponding to the peak in the cooling curve at \SI{e5}{K} (see
\S~3.2 of Paper~I),
since the thickness is set by the cooling length at higher speeds.  In
the case that the Alfv\'en speed is a significant fraction of the
sound speed, this will also tend to increase the thickness.  In
principle, the shell thickness can be measured observationally if the
source is sufficiently well resolved \citep{Kobulnicky:2017a},
although this is complicated by projection effects.  We return to the
issue of the shell thickness in the discussion below.

\subsubsection{Dust opacity}
\label{sec:dust-opacity}
The dust opacity will depend on the total dust-gas ratio and on the
composition and size distribution of the grains.  Our adopted value of
\SI{600}{cm^2.g^{-1}}, or \SI{1.3e-21}{cm^2.H^{-1}}, is appropriate
for average Galactic interstellar grains in the EUV and FUV (e.g.,
\citealp{Weingartner:2001a}), but there is ample evidence for
substantial spatial variations in grain extinction properties
\citep{Fitzpatrick:2007a}, both on Galactic scales
\citep{Schlafly:2016a} and within a single star forming region
\citep{Beitia-Antero:2017a}.  The properties of grains within
photoionized regions are very poorly constrained observationally
because the optical depth is generally much lower than in overlying
neutral material.  In the Orion Nebula, there is some evidence
\citep{Salgado:2016a} that the FUV dust opacity in the ionized gas may
be as low as \SI{90}{cm^2.g^{-1}}, although the uncertainties in this
estimate are large and different results are obtained in other
regions, such as W3(A) \citep{Salgado:2012a}.  It is even possible
that the FUV dust opacity may be larger than the ISM value if the
abundance of very small grains is enhanced through radiative torque
disruption of larger grains \citep{Hoang:2018a}.

\subsubsection{Shell gas temperature}
\label{sec:shell-gas-temp}
For bows around O~stars, the shell temperature should be close to the
photoionization equilibrium value of \(\approx \SI{e4}{K}\), since the
post-shock cooling length is short in ambient densities above
\SI{0.1}{cm^{-3}}
and the shell does not trap the ionization front for ambient densities
below \SI{e4}{cm^{-3}} (see Paper~I's \S\S~3.1 and 3.2 for details).
For B~stars, on the other hand, these two density limits move closer
together,
making it more likely that a bow will lie in a different temperature
regime.  The only source for which we have evidence that this occurs
is LP~Ori, as discussed in the next section.

\subsection{Special treatment of particular sources}
\label{sec:spec-treatm-part}

For two of the additional bows listed in Table~\ref{tab:observations},
we are forced to deviate from the default values for the shell
parameters.  For LP~Ori, the bow shell appears to be formed from
neutral gas \citep{ODell:2001c} and its relatively high \(\tau\) value is
more than sufficient to trap the weak ionizing photon output of a B3
star.  We therefore move its point in
Figure~\ref{fig:All-sources-eta-tau} downward by a factor of ten,
which could be a thermally supported neutral shell at \SI{2000}{K} or
a magnetically supported shell with
\(v\alfven \approx \SI{3}{km.s^{-1}}\)).  For the case of the Orion
Trapezium star \thD{}, we find that using the default parameters
results in a placement well inside the forbidden zone of
Figure~\ref{fig:All-sources-eta-tau} (indicated by fainter symbol).
For this object there is no reason to suspect anything but the usual
photoionized temperature of \SI{e4}{K}, but its placement could be
resolved either by decreasing the shell thickness, or decreasing the
UV dust opacity, or both. Given the moderate limb brightening seen in
the highest resolution images of the Ney--Allen nebula
\citep{Robberto:2005a, Smith:2005a}, the shell thickness is unlikely
to be less than half our default value.  But, if this were combined
with a factor 5 decrease in \(\kappa\), as suggested by
\citet{Salgado:2016a}, then this would be sufficient to move the
source up to the RBW line, or slightly above.

\subsection{Candidate radiation-supported bows}
\label{sec:cand-radi-supp}

Four sources are sufficiently close to the diagonal line
\(\eta\shell = 1.25 \tau\) in Figure~\ref{fig:All-sources-eta-tau}
that they should be treated as strong candidates for
radiation-supported bows. These are K18 sources~380 (HD~53367,
V750~Mon) and 407 (HD~93249 in Carina) plus \thD{} and LP~Ori.  Of the
four, source 380 is the only one that is also a candidate for
grain-gas decoupling.  Further details of the two K18 sources are
presented in Appendix~\ref{sec:notes-part-sourc}, where for source~380
we show that reducing both luminosity and distance by a factor of
roughly 2 with respect to the values used by K18 would provide a
better fit to the totality of observational data.  However, the ratio
\(\eta\shell /\tau\) is proportional to \(D / L_*\) so it would not be
affected by such an adjustment and the bow remains
radiation-supported.  On the other hand \(\eta\shell\) (proportional
to \(D / L_*^2\)) would increase by 2, making the classification as
dust wave candidate more marginal.

Three additional K18 sources (409, 410, 411) are within a factor of 3
of the radiation-supported line, so they too must be considered as
potential candidates, given our estimated systematic uncertainty in
the shell parameters (\S~\ref{sec:syst-uncert-due}).

\section{Mid-infrared grain emissivity}
\label{sec:grain-temp-emiss}

In preparation for our discussion of different wind mass-loss
diagnostic methods below, in this section we calculate the grain
emissivity predicted by models of dust heated by a nearby OB star.  We
use the same simulations that we employed in \S~4.2 of Paper~II,
which employ the plasma physics code Cloudy \citep{Ferland:2013a,
  Ferland:2017a}.  In summary, simulations of spherically symmetric,
steady-state, constant density \hii{} regions were carried out for
four different stellar types from B1.5 to O5 (Table~2 of Paper~II), a
range of gas densities from \SIrange{1}{e4}{cm^{-3}}, and using
Cloudy's default ``ISM'' graphite/silicate dust mixture with 10 size
bins from \SIrange{0.005}{0.25}{\um}.  

\begin{figure}
  \centering
  \includegraphics[width=\linewidth]{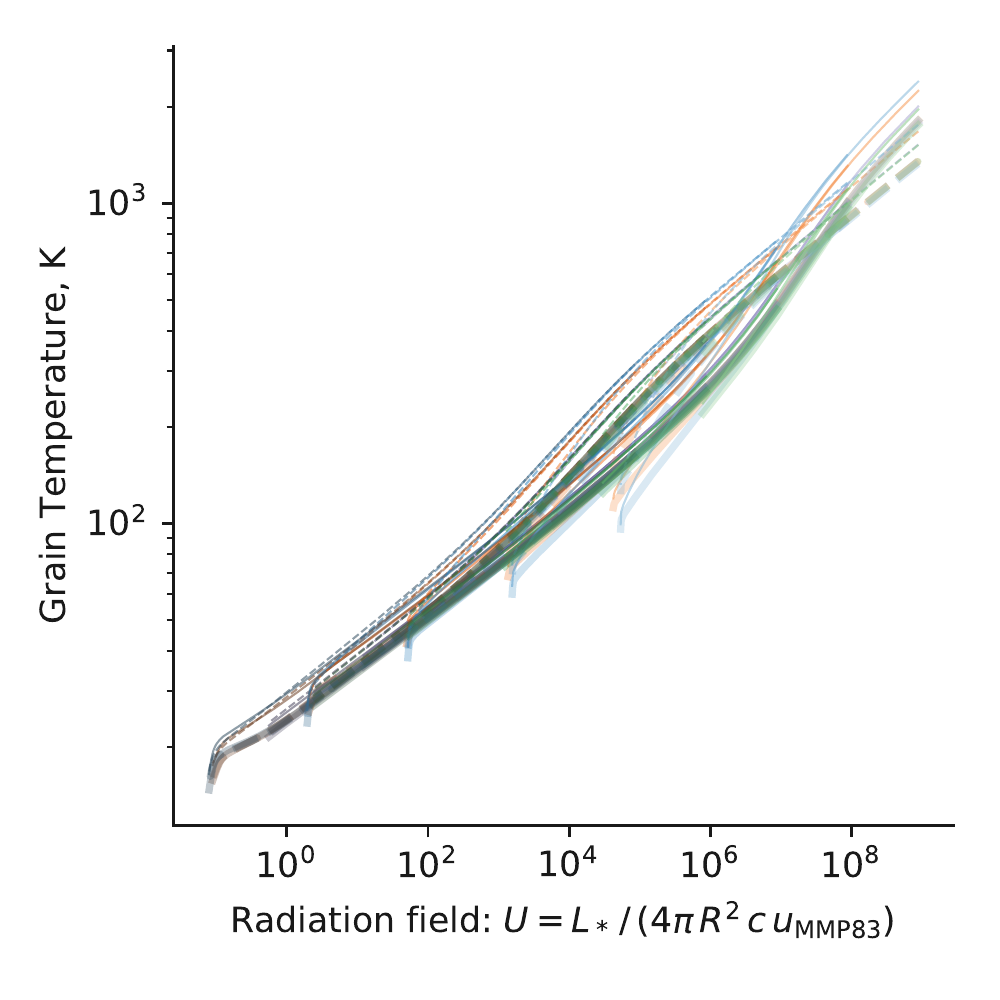}
  \caption{Grain temperature versus radiation field mean intensity,
    \(U\), in units of the interstellar radiation field in the solar
    neighborhood.  Line types and colors correspond to a variety of
    stellar spectral shapes, gas densities, and grain species.  Dashed
    lines show carbon grains, solid lines show silicate grains, with
    line thickness and transparency increasing with grain size.
    Stellar spectral types are O5\,V (blue), O9\,V (orange), B1.5\,V
    (purple), and B0.7\,Ia (green), with lighter shades denoting
    higher gas densities. }
  \label{fig:grain-T-vs-U}
\end{figure}

Figure~\ref{fig:grain-T-vs-U} shows equilibrium grain temperatures for
these Cloudy models as a function of the nominal energy density of the
radiation field, \(U = u / u\mmp \), where \(u = L / 4 \pi R^2 c\) and
\(u\mmp\) is the energy density of the interstellar radiation field
for \(\lambda < \SI{8}{\um}\) in the solar neighborhood
\citep{Mathis:1983a}:
\begin{equation}
  \label{eq:u-mmp83}
  u\mmp\,c = \SI{0.0217}{erg.s^{-1}.cm^{-2}} \ .
\end{equation}
The tight relationship seen in Figure~\ref{fig:grain-T-vs-U} between
\(T\) and \(U\) is evidence for the dominance of stellar radiative
heating, which we justify on theoretical grounds in
\S~\ref{sec:unimp-other-heat} below.  The variation about the mean
relation is mainly due to differences in grain size and composition,
with smaller grains and graphite grains being relatively hotter.  The
downward hooks seen on the left end of each simulation's individual
curve are due to the fact that our calculation of \(U\) does not
account for internal absorption, which starts to become important near
the ionization front.

\begin{figure}
  \centering
  \includegraphics[width=\linewidth]{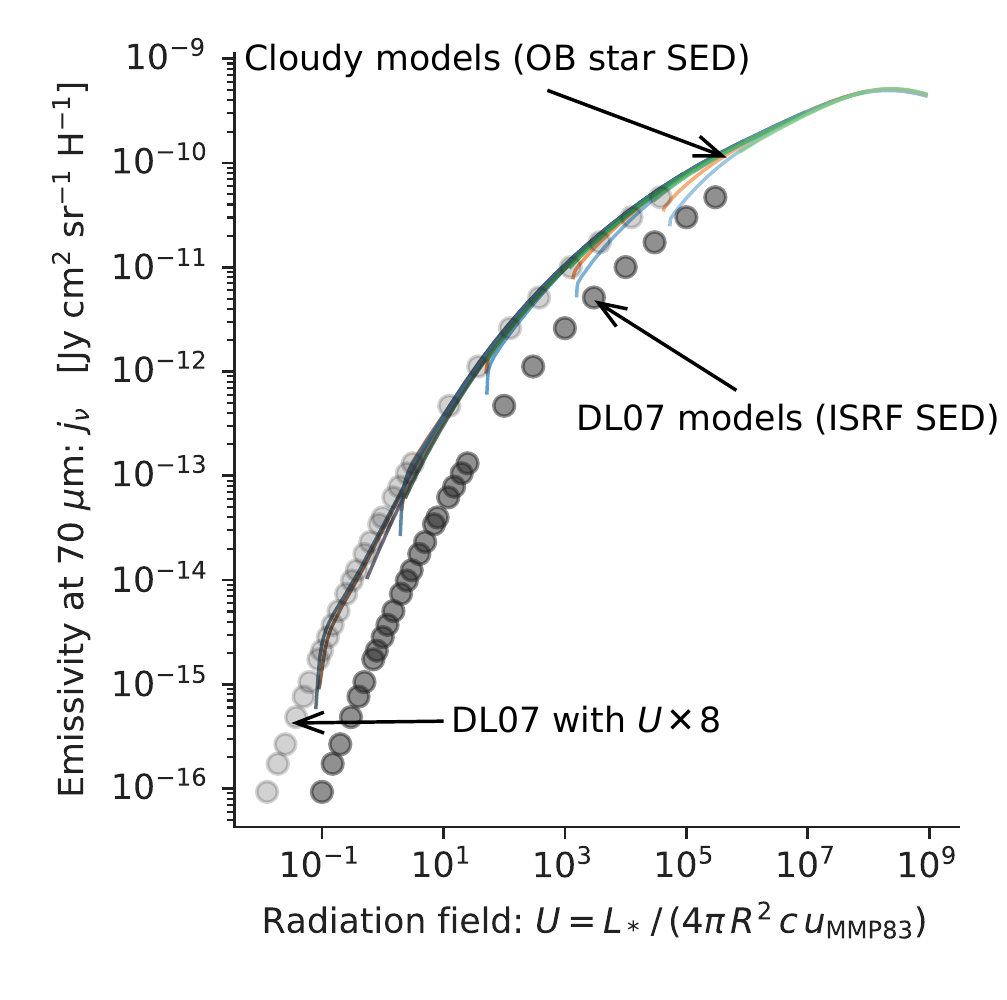}
  \caption{Grain emissivity at \SI{70}{\um} for all Cloudy models,
    with lines colored as in Fig.~\ref{fig:grain-T-vs-U}, but without
    the variation in line type and thickness since emissivity is
    integrated over all grain types and sizes.  For comparison, the
    emissivity from the grain models of \citet{Draine:2007a} are shown
    as dark gray symbols, which assume illumination by a scaled
    interstellar radiation field with a SED with a very different
    shape from that of an OB star, see Fig.~\ref{fig:sed-comparison}.
    The light gray symbols show the effect of using an 8 times higher
    \(U\) with the \citeauthor{Draine:2007a} models, which
    approximately compensates for this difference in SED.  }
  \label{fig:grain-j70}
\end{figure}

\begin{figure}
  \centering
  \includegraphics[width=\linewidth]{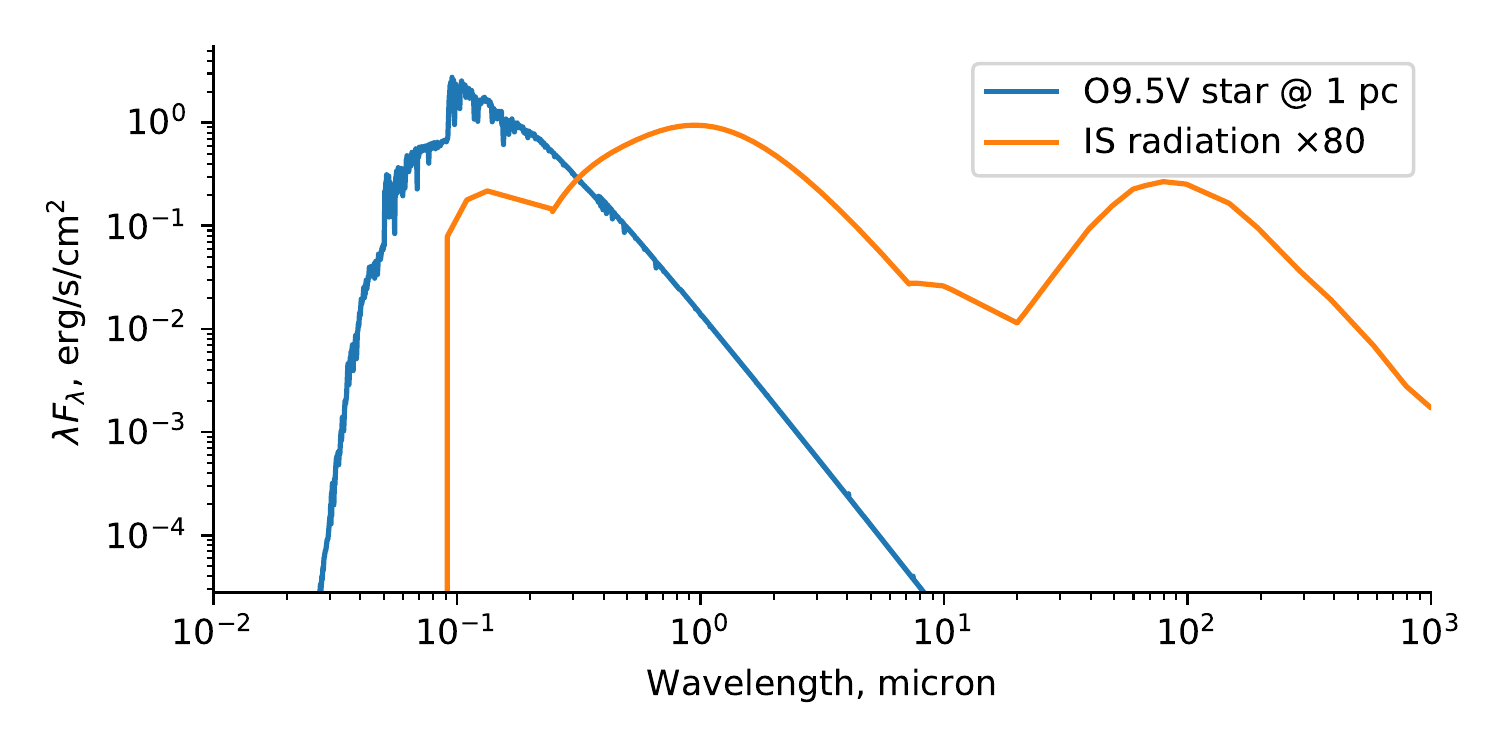}
  \caption{Comparison between the spectral energy distribution (SED)
    of a typical OB star (blue line) and the interstellar radiation
    field in the solar neighborhood (orange line).  The OB star is the
    \SI{20}{M_\odot} model from Table~1 of Paper~I and is plotted for
    a distance from the star of \SI{1}{pc}.  The interstellar SED is
    from \citet{Mathis:1983a} and is multiplied by \num{80} so that
    the total FUV-to-NIR flux is equal for the two SEDs.}
  \label{fig:sed-comparison}
\end{figure}

The grain emissivity at \SI{70}{\um} (Herschel PACS blue band) for the
Cloudy simulations (colored lines) is shown in
Figure~\ref{fig:grain-j70}, where it is compared with the same
quantity from the grain models (dark gray symbols) of
\citet{Draine:2007a}.  A clear difference is seen between the two sets
of models, but this is due almost entirely to a difference in the
assumed spectrum of the illuminating radiation, as illustrated in
Figure~\ref{fig:sed-comparison}.  \citet{Draine:2007a} use a SED that
is typical of the interstellar radiation field in the Galaxy, which is
dominated by an old stellar population, which peaks in the near
infrared, with only a small FUV contribution from younger stars (about
8\% of the total energy density).  This is very different from the OB
star SEDs, which are dominated by the FUV and EUV bands.  Since the
grain absorption opacity is substantially higher at UV wavelengths
than in the visible/IR (see Fig.~6 of Paper~II),
the effective grain heating efficiency of the OB star SED is
correspondingly higher.  The light gray symbols show the effect on the
\citet{Draine:2007a} models of multiplying the radiation field by a
factor of \num{8} in order to offset this difference in efficiency,
which can be seen to bring them into close agreement with the Cloudy
models.  A further difference is that the \citet{Draine:2007a} model
includes small PAH particles, which we do not include in our Cloudy
models, since they are believed to be largely absent in photoionized
regions \citep{Giard:1994a, Lebouteiller:2011a}.  However, this only
effects the emissivity at shorter mid-infrared wavelengths
\(< \SI{20}{\um}\).

In terms of the characteristic parameters introduced in
\S~\ref{sec:energy-trapp-vers} the dimensionless radiation field
becomes
\begin{equation}
  \label{eq:U-from-L4-and-Rpc}
  U = 14.7\, L_4\, R_{\text{pc}}^{-2} \ ,
\end{equation}
or, alternatively, it can be expressed in terms of the ambient stream as
\begin{equation}
  \label{eq:U-from-ambient}
  U = 3.01 \, n_\infty \, v_{10}^2 / x^2 \ , 
\end{equation}
where \(x = R_0/R_*\) is given by Paper~I's equation~(12).
It can also be related to the radiation parameter \(\Xi\), defined in
Paper~II's equation~(23), as
\begin{equation}
  \label{eq:U-vs-Xi}
  U = 3.82 \, n T_4 \, \Xi \ .
\end{equation}
A common alternative approach to scaling the radiation field (see
\citealp{Tielens:1985a} and citations thereof) is to normalize in the
FUV band (\SIrange{0.0912}{0.24}{\um}), where the local interstellar
value is known as the Habing flux \citep{Habing:1968a}:
\begin{equation}
  \label{eq:Habing-flux}
  F\Hab = \SI{0.0016}{erg.s^{-1}.cm^{-2}} \ .
\end{equation}
The resultant dimensionless flux is often denoted by \(G_0\), and the
relationship between \(G_0\) and \(U\) depends on the fraction
\(f_{\text{fuv}}\) of the stellar luminosity that is emitted in the
FUV band:
\begin{equation}
  \label{eq:G-vs-U}
  G_0 = f_{\text{fuv}} \frac{u\mmp\, c}{F\Hab} \,U = (\text{\numrange{6}{10}}) \,U \ ,
\end{equation}
where we give the range corresponding to early O (\(f_{\text{fuv}} \approx 0.4\)) to early B (\(f_{\text{fuv}} \approx 0.7\)) stars.


\subsection{Unimportance of other heating mechanisms}
\label{sec:unimp-other-heat}
The grain temperature in bows around OB stars is determined
principally by the steady-state equilibrium between the absorption of
stellar UV radiation (heating) and the thermal emission of infrared
radiation (cooling).  Other processes such as single-photon stochastic
heating, Lyman~\(\alpha\) line radiation, and post-shock collisional
heating can dominate in other contexts, but these are generally
unimportant for circumstellar bows, as we now demonstrate.

\subsubsection{Stochastic single-photon heating}
\label{sec:stoch-single-phot}
When the radiation field is sufficiently dilute, then a grain that
absorbs a photon has sufficient time to radiate all that energy away
before it absorbs another photon \citep{Duley:1973a}.  In this case,
the emitted infrared spectrum for \(\lambda < \SI{50}{\um}\) becomes
relatively insensitive of the energy density of the incident radiation
\citep{Draine:2001a}.  However, this is most important for the very
smallest grains.  From equation~(47) of \citet{Draine:2001a}, one
finds that grains with sizes larger than
\(a = \SI{0.005}{\um} = \SI{5}{nm}\) (the smallest size included in
our Cloudy models) should be close to thermal equilibrium for
\(U > 30\), which is small compared with typical bow shock values
(\(U = \text{\numrange{e3}{e6}}\)).  As mentioned above, PAHs are not
expected to be present in the interior of \hii{} regions.
\citealp{Desert:1990a} found them to be strongly depleted for
\(U > 100\) around O stars.  However, other types of ultra-small
grains, down to sub-nm sizes \citep{Xie:2018a} may be present in bows,
and stochastic heating \emph{would} be important for grains with
\(a = \SI{1}{nm}\) if \(U < \num{e5}\).  Note, however that grains
smaller than \SI{0.6}{nm} would be destroyed by sublimation after
absorbing a single He-ionizing photon.

\subsubsection{Lyman \(\alpha\) heating}

On the scale of an entire \hii{} region, the dust heating is typically
dominated by Lyman \(\alpha\) hydrogen recombination line photons, which
are trapped by resonant scattering (e.g., \citealp{Spitzer:1978a}
\S~9.1b).  However, this is no longer true on the much smaller scale
of typical bow shocks.  An upper limit on the Lyman \(\alpha\) energy
density can be found by assuming all line photons are ultimately
destroyed by dust absorption rather than escaping in the line wings
(e.g., \citealp{Henney:1998b}), which yields
\begin{equation}
  \label{eq:U-Lya}
  U\Lya \approx 0.1 n / \kappa_{600} \ .
\end{equation}
This can be combined with equation~\eqref{eq:U-from-ambient} to give
the ratio of Lyman \(\alpha\) to direct stellar radiation as
\begin{equation}
  \label{eq:Lya-over-stellar}
  \frac{U\Lya}{U} \approx 0.03 \frac{x^2}{v_{10}^2 \kappa_{600}} \ .
\end{equation}
Taking the most favorable parameters imaginable of a slow stream
(\(v_{10} = 2\)), very strong wind (\(x \approx 1\)), and reduced dust
opacity (\(\kappa_{600} = 0.1\)) gives a Lyman \(\alpha\) contribution of only
10\% of the stellar radiative energy density.  In any other
circumstances, the fraction would be even lower.

\subsubsection{Shock heating}
\newcommand\kin{\ensuremath{_{\text{kin}}}}
The outer shock thermalizes the kinetic energy of the ambient stream,
which may in principle contribute to the infrared emission of the bow.
In order for this process to be competitive, the following three
conditions must all hold:
\begin{enumerate}[1.]
\item The post-shock gas must radiate efficiently with a cooling
  length less than the bow size, see \S~3.2 of Paper~I.\@
  This is satisfied for all but the lowest densities (Paper~I's
  Fig.~2).
\item A significant fraction of the shock energy must be radiated by
  dust.  This requires that the post-shock temperature be greater than
  \SI{e6}{K}, which requires a stream velocity
  \(v_\infty > \SI{200}{km.s^{-1}}\) \citep{Draine:1981a}.  This also
  coincides with the range of shock velocities where the smaller
  grains will start to be destroyed by sputtering in the post-shock
  gas.
\item The kinetic energy flux through the shock must be significant,
  compared with the fraction of the stellar radiation flux that is
  absorbed and reprocessed by the bow shell.
\end{enumerate}
It turns out that the third condition is the most stringent, so we
will consider it in detail.  The kinetic energy flux through the outer shock for an ambient stream of density \(\rho_\infty\) and velocity \(v_\infty\) is
\begin{equation}
  \label{eq:Fkin}
  F\ke = \tfrac12 \rho_\infty v_\infty^3 = \tfrac12 P\shell v_\infty \ , 
\end{equation}
while the stellar radiative energy flux absorbed by the shell is
\begin{equation}
  \label{eq:Ftrap}
  F\trap \approx \tau L / 4 \pi R_0^2 \ ,
\end{equation}
assuming an absorption optical depth \(\tau \ll 1\). The shell
pressure in the WBS case can be equated to the ram pressure of the
internal stellar wind (see
\S~2.1 of Paper~I), so that the ratio of the two energy fluxes is
\begin{equation}
  \label{eq:F-ratio-shock}
  \frac{F\ke}{F\trap} = \frac12 \frac{\eta\wind}{\tau} \frac{v_\infty}{c} \ .
\end{equation}
An upper limit to the stellar wind momentum efficiency \(\eta\wind\) is
the shell momentum efficiency \(\eta\shell\) that is derived
observationally in \S~\ref{sec:energy-trapp-vers}, where it is found
that \(\eta\shell / \tau < 30\) for all sources considered.  Therefore, for
a stream velocity \(v_\infty = \SI{200}{km.s^{-1}}\), we have
\(F\ke/F\trap < 0.01\) and the shock-excited dust emission is still
negligible.  Only in stars with \(v_\infty > \SI{1000}{km.s^{-1}}\) would
the shock emission start to be significant, and such hyper-velocity
stars \citep{Brown:2015a} do not show detectable bow shocks.

So far, we have only considered the outer shock, but the inner shock
that decelerates the stellar wind will have a velocity of
\SIrange{1000}{3000}{km.s^{-1}} and therefore might have a significant
kinetic energy flux by eq.~\eqref{eq:F-ratio-shock}.  However, the
stellar wind from hot stars will be free of dust,\footnote{%
  With the exception of Wolf-Rayet colliding wind binary systems
  \citep{Tuthill:1999a, Callingham:2019a}.} %
so that it would be necessary for the stellar wind protons to cross
the contact/tangential discontinuity and deposit their energy in the
dusty plasma of the shocked ambient stream in order for this source of
energy to contribute to the grain emission.  This is not possible
because the Larmor radius (see \S~5 of Paper~II)
of a \SI{3000}{km.s^{-1}} proton in a \SI{1}{\micro G} field is only
\SI{3e10}{cm}, which is millions of times smaller than typical bow
sizes.  The magnetic field in the outer shell is unlikely to be
smaller than \(\approx n^{1/2} \si{\micro G}\), given that Alfvén speeds of
\SI{2}{km.s^{-1}} are typical of photoionized regions \citep{Arthur:2011a, Planck-Collaboration:2016c}
and if the density were much lower than
\SI{1}{cm^{-3}}, then the scale of the bow would be commensurately
larger anyway.  Three-dimensional MHD simulations of bow shocks
\citep{Katushkina:2017a, Gvaramadze:2018a} show that the magnetic
field lines are always oriented parallel to the shell, so that high
energy particles from the stellar wind would be efficiently reflected
in a very thin layer and cannot contribute to grain heating.  For the
same reason, heat conduction by electrons across the contact
discontinuity is also greatly suppressed \citep{Meyer:2017a}.


\begin{figure*}
  \centering
  \includegraphics[width=\linewidth]{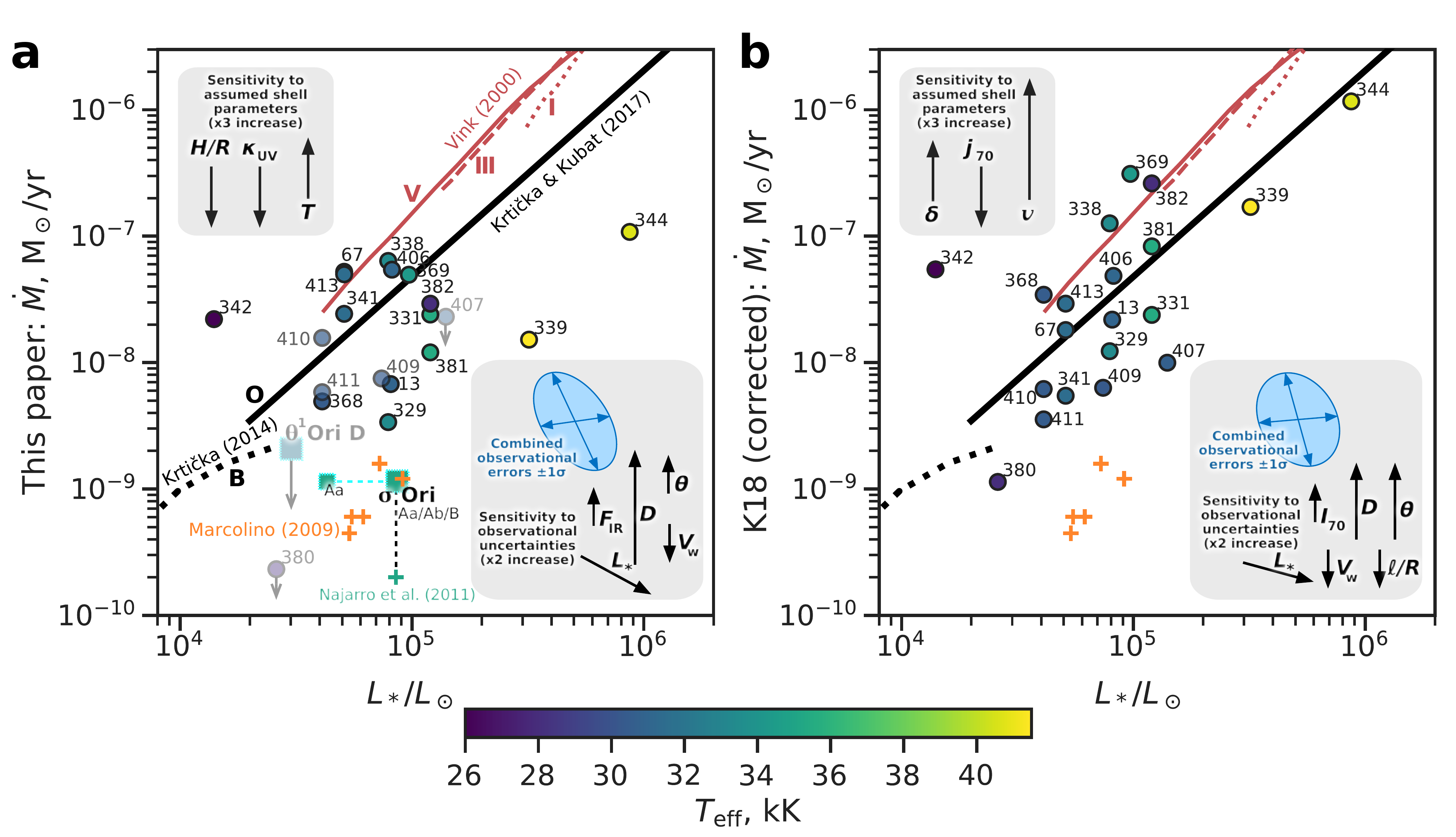}
  \caption{Wind mass-loss rates as a function of stellar luminosity,
    derived from (a)~our trapped energy/momentum method and (b)~the
    grain emissivity method of \citet{Kobulnicky:2018a}, with
    corrections as described in our \S~\ref{app:bow-shock-data}.
    Circle symbols show the sources from K18, colored according to the
    stellar effective temperature (see key at far right). In panel a,
    squares show two of our additional sources
    (Tab.~\ref{tab:observations}). Upper limits to the mass loss are
    given for sources that lie close to the radiation-supported line
    in Fig.~\ref{fig:All-sources-eta-tau}, represented by faint
    symbols and downward-pointing arrows.  Sources that lie less than
    a factor of three above the line have enhanced downward
    uncertainties and are also shown by slightly fainter symbols.  For
    \(\sigma\)~Ori, the large symbol corresponds to the sum of the
    luminosities of the triple OB system Aa, Ab, and B
    \citep{Simon-Diaz:2015a}, while the small symbol corresponds to
    the luminosity of only the most massive component Aa.  Lines show
    the predictions of stellar wind models: red lines are the commonly
    used recipes from \citet{Vink:2000a} for dwarfs (solid), giants
    (dashed), and supergiants (dotted), while black lines show
    eq.~(11) of \citet{Krticka:2017a} for O~stars (solid) and models
    of \citet{Krticka:2014a} for B stars.  Orange plus symbols show
    mass-loss measurements from NUV lines for weak-wind O~dwarfs
    \citep{Marcolino:2009a}, while the green plus symbol shows the
    measurement from infrared H recombination lines for \(\sigma\)~Ori
    \citep{Najarro:2011a}.  Boxes show the sensitivity of the results
    to observational uncertainties (lower right) and assumed shell
    parameters (upper left).}
  \label{fig:mass-loss-vs-luminosity}
\end{figure*}

\section{Stellar wind mass-loss rates}
\label{sec:stellar-wind-mass}

Various methods have been proposed to derive stellar wind mass-loss
rates from observations of stellar bow shocks \citetext{for example,
  \citealp{Kobulnicky:2010a, Gvaramadze:2012a, Kobulnicky:2018a}}.  In
this section, we will show how the \(\tau\)--\(\eta\shell\) diagram can be
used to derive the mass-loss rate for bows in the wind-supported
regime.  We then compare our method with the method used by
\citet{Kobulnicky:2018a}, which is a refinement of that originally
proposed in \citet{Kobulnicky:2010a}.  The method used by
\citet{Gvaramadze:2012a} uses combined measurements of the bow shock
and the surrounding \hii{} region.  It has the advantage of depending
on fewer free parameters than the other methods, but can only be used
in the case of isolated stars, whereas the majority of the sources
considered here are in cluster environments.

\subsection{Mass loss determination from the \boldmath \(\tau\)--\(\eta\shell\) diagram}
\label{sec:mass-loss-determ}

Taking into account the support from stellar wind ram pressure and the
absorbed stellar radiation pressure, the pressure of the bow shell can
be written
\begin{equation}
  \label{eq:eta-shell-components}
  \eta\shell = \eta\wind +  \left(  1 - e^{-\Qp \tau / Q_{\text{abs}}}\right) \ ,
\end{equation}
which is valid in all 3 regimes: WBS, RBW, and RBS.\@ The wind
momentum efficiency is defined in terms of the stellar wind parameters
(eq.~[13] of Paper~I):
\begin{equation}
  \label{eq:wind-eta-typical}
  \eta\wind = \num{0.495} \,\dot{M}_{-7} \,V_3  \,L_4^{-1} \ .
\end{equation}
Therefore, assuming \(\Qp/Q_{\text{abs}} = 1.25\) as in
\S~\ref{sec:energy-trapp-vers}, the mass-loss rate can be estimated as
\begin{equation}
  \label{eq:mdot-tau-eta}
  \dot{M}_{-7} = 2.02 \, L_4 \, V_3^{-1} \,
  \left[ \eta\shell - \left(  1 - e^{-1.25 \tau} \right) \right]  \ .
\end{equation}
In the wind-supported WBS regime
(\(\tau \ll \eta\wind \approx \eta\shell\)), the second term in the brackets is
negligible, and we can use equations~(\ref{eq:tau-empirical},
\ref{eq:eta-shell}) to write
\begin{equation}
  \label{eq:mdot-eta-wbs}
  \dot{M}_{-7} \approx 990 \frac{R\pc \, T_4 \, L_{\text{IR},4}}
  {L_4 \, V_3 \, \kappa_{600} \, h_{1/4}} \ .
\end{equation}
On the other hand, if \(\eta\shell\) does not greatly exceed \(\tau\), then
the full equation~\eqref{eq:mdot-tau-eta} should be used, although the
uncertainties in \(\eta\shell\) and \(\tau\), combined with the partial
cancellation of two similar-sized terms, mean that the mass loss will
be poorly constrained in such cases.

Resulting mass-loss rates are shown as a function of stellar
luminosity in Figure~\ref{fig:mass-loss-vs-luminosity}a for the K18
and Orion sources.  As in Figure~\ref{fig:All-sources-eta-tau} we show
separately the uncertainties in the measurements due to observational
errors (lower right box) and systematic model uncertainties (upper
left box).  Typical observational uncertainties are combined into an
error ellipse using the techniques of
Appendix~\ref{sec:comb-uncert-covar}, which is shown in light blue.
Sources that are close to the RBW line in the \(\tau\)--\(\eta\shell\)
diagram yield only upper limits to the mass-loss rate and these are
shown as faint symbols in the figure.

Our results are compared with the commonly used mass-loss recipes of
\citet{Vink:2000a} (red lines) and more recent whole-atmosphere
simulations of wind formation in B~stars \citep{Krticka:2014a} and
O~stars \citep{Krticka:2017a}.  Additionally, we show observational
mass loss determinations for weak wind O~dwarfs
\citep{Marcolino:2009a} as orange plus symbols.  It can be seen that
the majority of the K18 sources fall below the \citeauthor{Vink:2000a}
line and are more consistent with the \citet{Krticka:2017a} models.

\subsection{Mass loss determination method of \protect\citet{Kobulnicky:2018a}}
\label{app:bow-shock-data}



K18 derive mass-loss rates for their sources using a method that is
different from the one that we employ above.  Both methods are based
on determining the stellar wind ram pressure that supports the bow
shell, but K18 do so via the following steps:
\begin{enumerate}[K1.]
\item \label{K1} The line-of-sight mass column through the shell is
  calculated by combining the peak surface brightness at \SI{70}{\um},
  \(S_{70}\), with a theoretical emissivity per nucleon,
  \(j_{70}(U)\), from \citet{Draine:2007a}:
  \(\Sigma\LOS = S_{70} / \bar{m} j_{70}(U)\).  This depends on
  knowledge of the stellar radiation field at the shell:
  \(U \propto L_* / R_0^2\).
\item \label{K2} The shell density is found from the line-of-sight
  mass column using an observationally determined ``chord diameter'',
  \(\ell\), which is assumed to be equal to the depth along the line
  of sight: \(\rho\shell = \Sigma\LOS / \ell\).
\item \label{K3} The internal ram pressure is equated to the external
  ram pressure, which is found by assuming a stream velocity of
  \SI{30}{km.s^{-1}} and a compression factor of 4 across the outer
  shock:
  \(P_{\text{stream}} = 0.25 \rho\shell \times
  (\SI{30}{km.s^{-1}})^2\).
\end{enumerate}
There are clear parallels but also differences between steps
K\ref{K1}--K\ref{K3} and our own steps P\ref{P1}--P\ref{P3}.  Our
step~P\ref{P1} depends on the total observed infrared flux of the bow
combined with an assumption about the grain opacity at ultraviolet
wavelengths, while step~K\ref{K1} depends on the peak brightness at a
single wavelength combined with an assumption about the grain
emissivity at infrared wavelengths.  Our step~P\ref{P2} requires an
assumption about the relative thickness of the shell, while
step~K\ref{K2} is more directly tied to observations.
On the other hand, step~K\ref{K3} makes a roughly equivalent
assumption about the shock compression factor,\footnote{%
  In reality, the compression factor may be larger or smaller than 4,
  depending on the efficiency of the post-shock cooling (see
  \S~3.2 of Paper~I).  For instance, for \(v = \SI{30}{km.s^{-1}}\) as
  assumed by K18 and \(T = \SI{e4}{K}\), one has a Mach number of
  \(\M_0 = 2.63\) and a compression factor of 2.8 for a non-radiative
  shock (by
  eq.~[27] of Paper~I) or a factor of
  \(\M_0^2 =6.9\) for a strongly radiative one.} %
and a further assumption about the stream velocity. These assumptions
are not necessary for our step~P\ref{P3}, but we do need to assume a
value for the shell gas temperature.

In principle, both methods are valid and their different assumptions
and dependencies on observed quantities and auxiliary parameters
provide an important cross check on one another.  However, as
explained in detail in \S~\ref{sec:grain-temp-emiss}, the \(j_\nu(U)\)
relation depends on the shape of the illuminating SED, which means
that the \citet{Draine:2007a} models require modification when applied
to grains around OB stars.  A further discrepancy arises due to an
interpolation error in K18, which resulted in values of \(j_{70}\)
being overestimated by about a factor of 2 for sources with weak
radiation fields (H.~Kobulnicky, priv.~comm.).



\begin{figure}
  \centering
  \includegraphics[width=\linewidth]{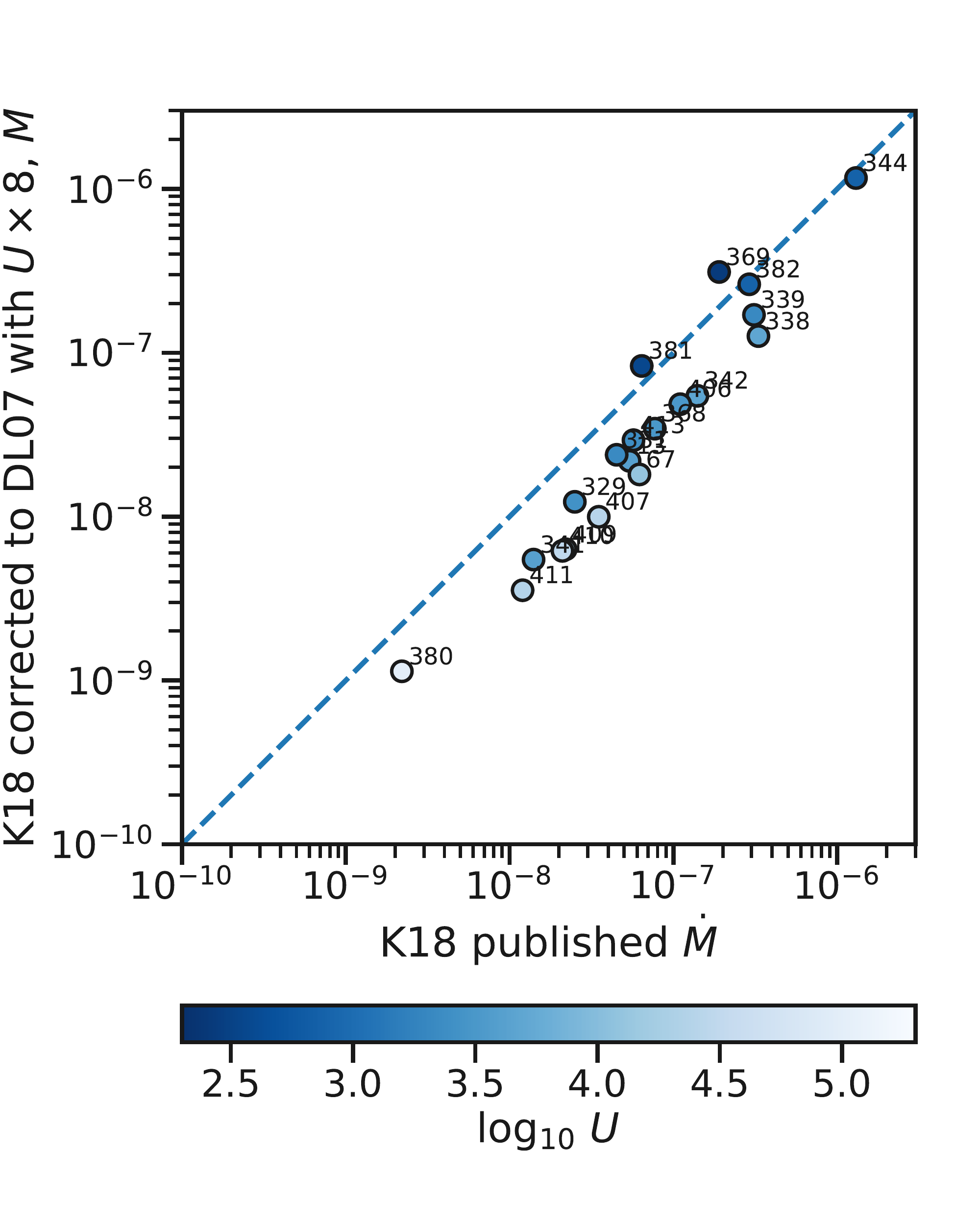}
  \caption{Effects on mass-loss determination of correcting the K18
    emissivities.  The mass-loss rates from Table~2 of K18 are shown
    on the \(x\) axis, while the corrected values are shown on the
    \(y\) axis.  Symbols are color coded by the strength of the
    radiation field, \(U\). The corrected mass-loss rates are
    predominantly lower by a factor of roughly 2.}
  \label{fig:k18-mdot-corrected-emissivity}
\end{figure}

After correcting the \SI{70}{\um} emissivities in this way, we
re-derive the mass loss rates, following the same steps as in K18,
which are then used in Figure~\ref{fig:mass-loss-vs-luminosity}b
above.  The difference between these corrected mass-loss rates and
those published in K18 is shown in
Figure~\ref{fig:k18-mdot-corrected-emissivity}.  It can be seen that
sources with \(U \approx \num{e3}\) (darker shading) are relatively
unaffected but that sources with stronger radiation fields (lighter
shading) have their mass-loss increasingly reduced.
The average reduction is by a factor of about two.




\section{Discussion}
\label{sec:discussion}

In this section we discuss various issues related to our results,
beginning in \S~\ref{sec:evolution-grain-size} with a consideration of
how the optical properties of the grain population in bow shocks might
be affected by the extreme radiation environment.  This is followed by
examination of various sub-groups of sources that are unusual in some
way: those where the two mass-loss methods give discrepant results
(\S~\ref{sec:underst-diff-betw}), those that may be
radiation-supported (\S~\ref{sec:comm-prop-cand}), and one source that
shows a very large apparent mass-loss rate for its luminosity
(\S~\ref{sec:anomalous-b-star}).

\subsection{Evolution of grain size}
\label{sec:evolution-grain-size}

\begin{figure}
  \centering
  \includegraphics[width=\linewidth]{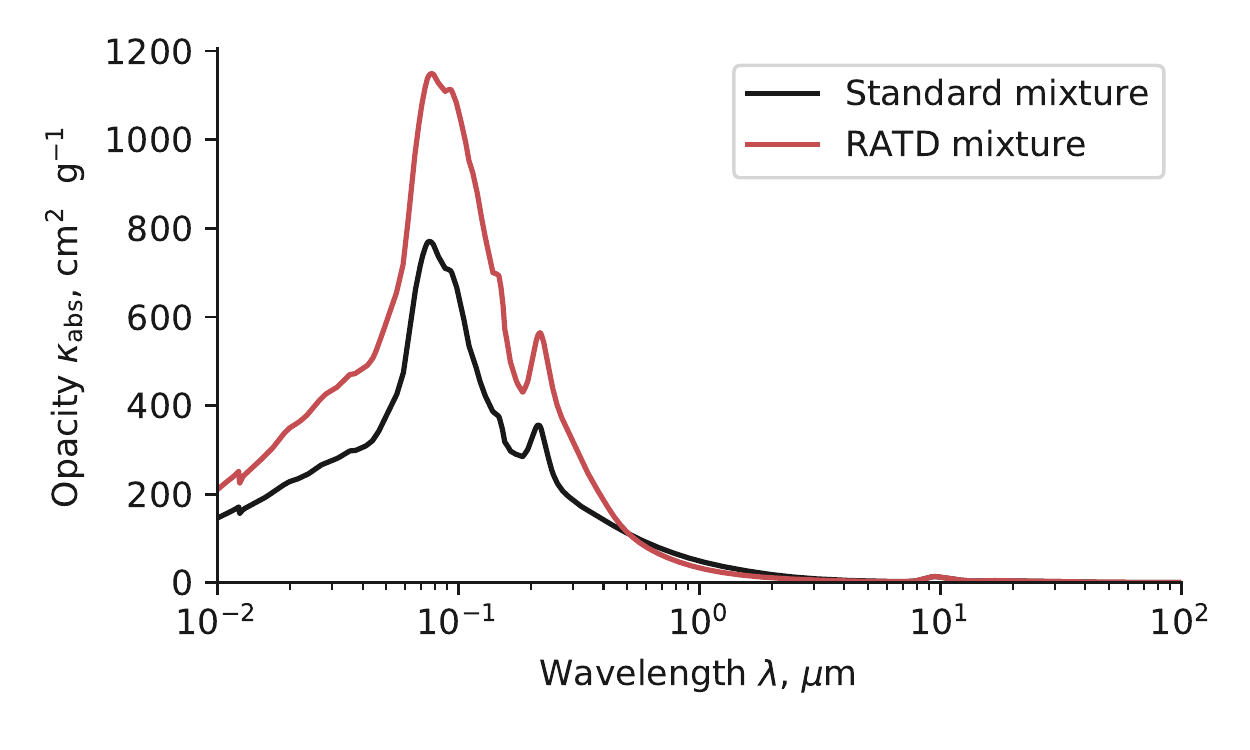}
  \caption{Change in dust opacity due to Radiative Torque Disruption
    of large grains.  Black line shows the total absorption opacity of
    Cloudy's standard ISM dust mixture.  The red line shows the result
    of simulating the RATD process by removing the grains larger than
    \SI{0.04}{\um} and distributing their mass among smaller grain
    sizes.}
  \label{fig:ratd}
\end{figure}

The properties of dust grains in bow shocks might be significantly
different from those in the general interstellar medium, particularly
because of the strong radiation field to which they are exposed, which
has effects at both the small and large ends of the grain size
distribution.  At the small end, sub-nanometer sized particles, such
as PAHs can be destroyed by hard EUV photons
\citep{Lebouteiller:2007a, Lebouteiller:2011a}.  Larger grains, on the
other hand, are most vulnerable to Radiative Torque Disruption
\citep[RATD;][]{Hoang:2018a}, see discussion in \S~6.4 of Paper~II.\@
This is important for our diagnostics because it would tend to reduce
the average grain size, while maintaining the same total grain mass,
which would have the effect of increasing the ultraviolet opacity,
while leaving the mid-infrared opacity unchanged.\footnote{The
  mid-infrared emissivity, on the other hand, would also increase
  since smaller grains tend to have higher radiative equilibrium
  temperatures.} %
We do not know the exact size distribution of fragments that would
result from the RATD process, but we can estimate its effect by
assuming that all grains with \(a > \SI{0.05}{\um}\) are transformed
into grains with \(a = \text{\SIrange{0.025}{0.05}{\um}}\).  We
implement this crudely in the Cloudy grain model
(\S~\ref{sec:grain-temp-emiss}) by setting to zero the abundance of
graphite and silicate grains in size bins 7, 8, 9, 10, while at the
same time distributing their mass equally between size bins 5, and 6,
which increases the abundance in those bins by factors of 5.19 and
4.45, respectively.  The results are shown in Figure~\ref{fig:ratd},
which shows the wavelength-dependent absorption opacity \(\kappa\) for
both the standard ISM grain mixture and this RATD-modified mixture.
As expected, the UV opacity is increased in the RATD mixture, but only
by about 50\%, whereas the near-infrared opacity is decreased and the
visual extinction becomes much steeper, which would correspond to a
small total-to-selective extinction ratio of \(R_V < 3\).

This modest increase in opacity from RATD grain processing is well
within the systematic uncertainties that we have been assuming and so
does not invalidate the results of the previous sections.  If the
larger grains were instead to break up into many small fragments, the
effect would be much larger.  If we repeat the above exercise, but
assuming the fragment size is \(< \SI{0.01}{\um}\), then we find an
increase by a factor of five in the UV opacity, but we do not feel
that this is realistic. Studies of the break-up of fast-spinning small
asteroids \citep{Hirabayashi:2015a, Zhang:2018a} show that when the
tensile strength dominates over self-gravity, then stresses are
highest in the center of the body, leading to break up into a small
number of similarly sized pieces, as observed in asteroid P/2013~R3
\citep{Jewitt:2014a}.  In the case of dust grains, the tensile
strength and angular velocity are both far higher than in the asteroid
case, but we expect a similar behavior so that the fragment size
should be predominantly within a factor of about two below the
critical size given in equation~(28) of \citet{Hoang:2018a}, as we
assumed in Figure~\ref{fig:ratd}.

\subsection{Why do results of the two mass-loss methods differ?}
\label{sec:underst-diff-betw}

\begin{figure}
  \centering
  \includegraphics[width=\linewidth]{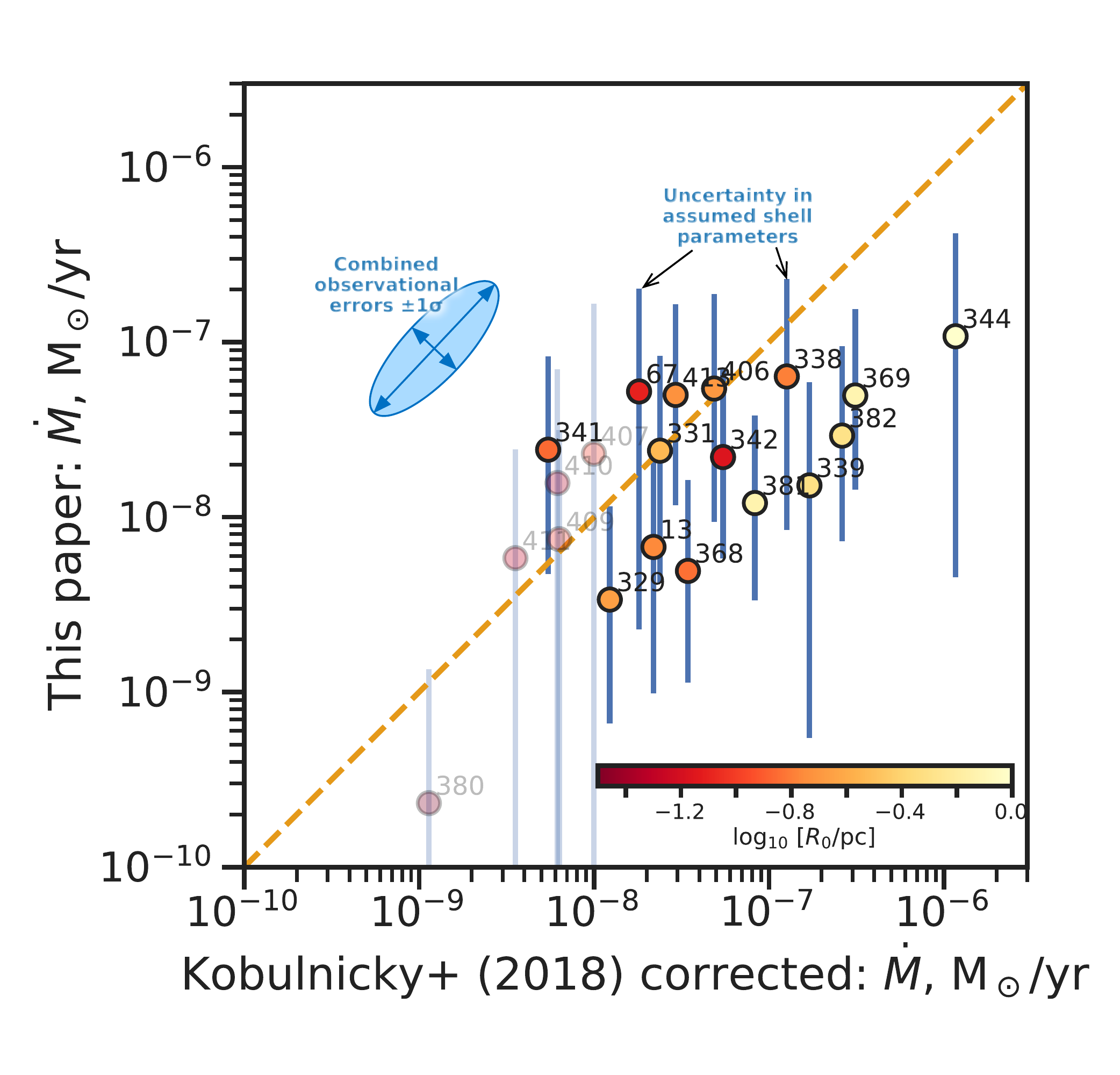}
  \caption{Comparison of the two mass-loss methods: K18 corrected
    method (\(x\) axis) versus our method (\(y\) axis).  Error bars on
    the \(y\) axis correspond to a factor-three uncertainty in
    \(\eta\shell\).  Sources for which these error bars overlap with
    the RBW zone are only upper limits for the wind mass-loss rate,
    and are indicated by faint symbols.}
  \label{fig:mass-loss-comparison}
\end{figure}

\begin{figure*}
  \centering
  \includegraphics[width=\linewidth]{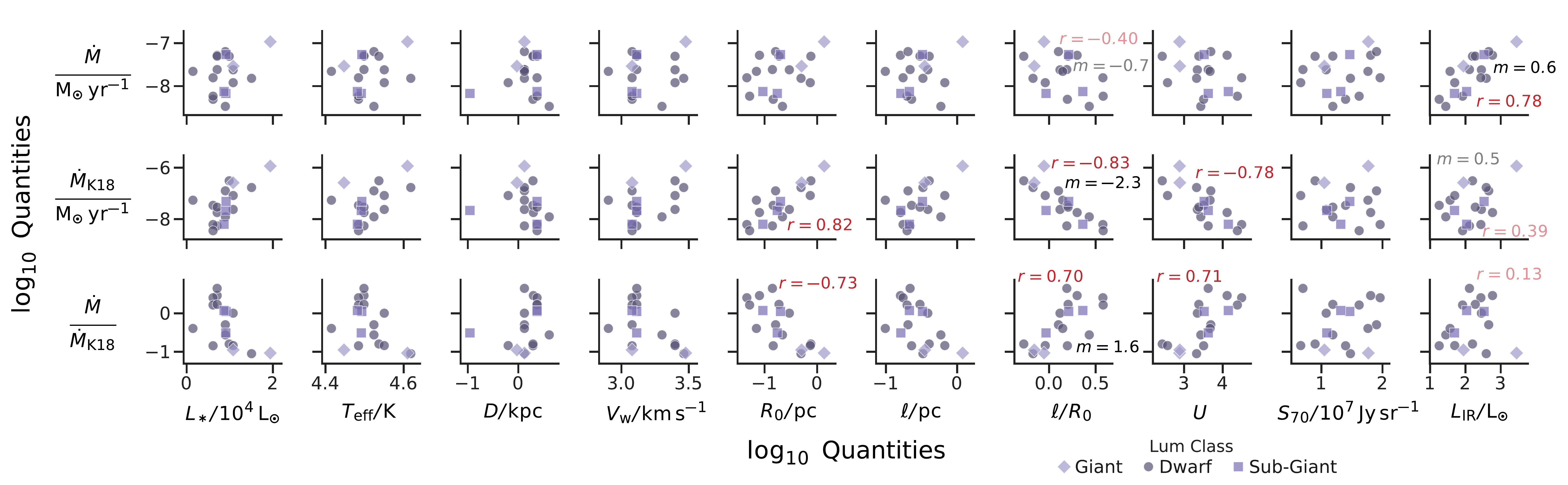}
  \caption{Correlations of derived mass loss rates with star and bow
    parameters. First row is for mass-loss rate derived using the
    method of this paper (\S~\ref{sec:mass-loss-determ}).  Second row
    is for mass-loss rate derived using the corrected K18 method
    (\S~\ref{app:bow-shock-data}).  Third row is the ratio of these
    two methods.  Points show 18 of the 20 K18 sources, omitting the
    two strongest candidates for radiation support.  The correlation
    coefficient, \(r\), and linear regression slope, \(m\), are shown
    for selected pairs of interest (fainter text indicates weaker
    correlations).  These were calculated using the Python library
    function \texttt{scipy.stats.linregress}.}
  \label{fig:correlations}
\end{figure*}

\begin{figure}
  \centering
  \includegraphics[width=\linewidth]{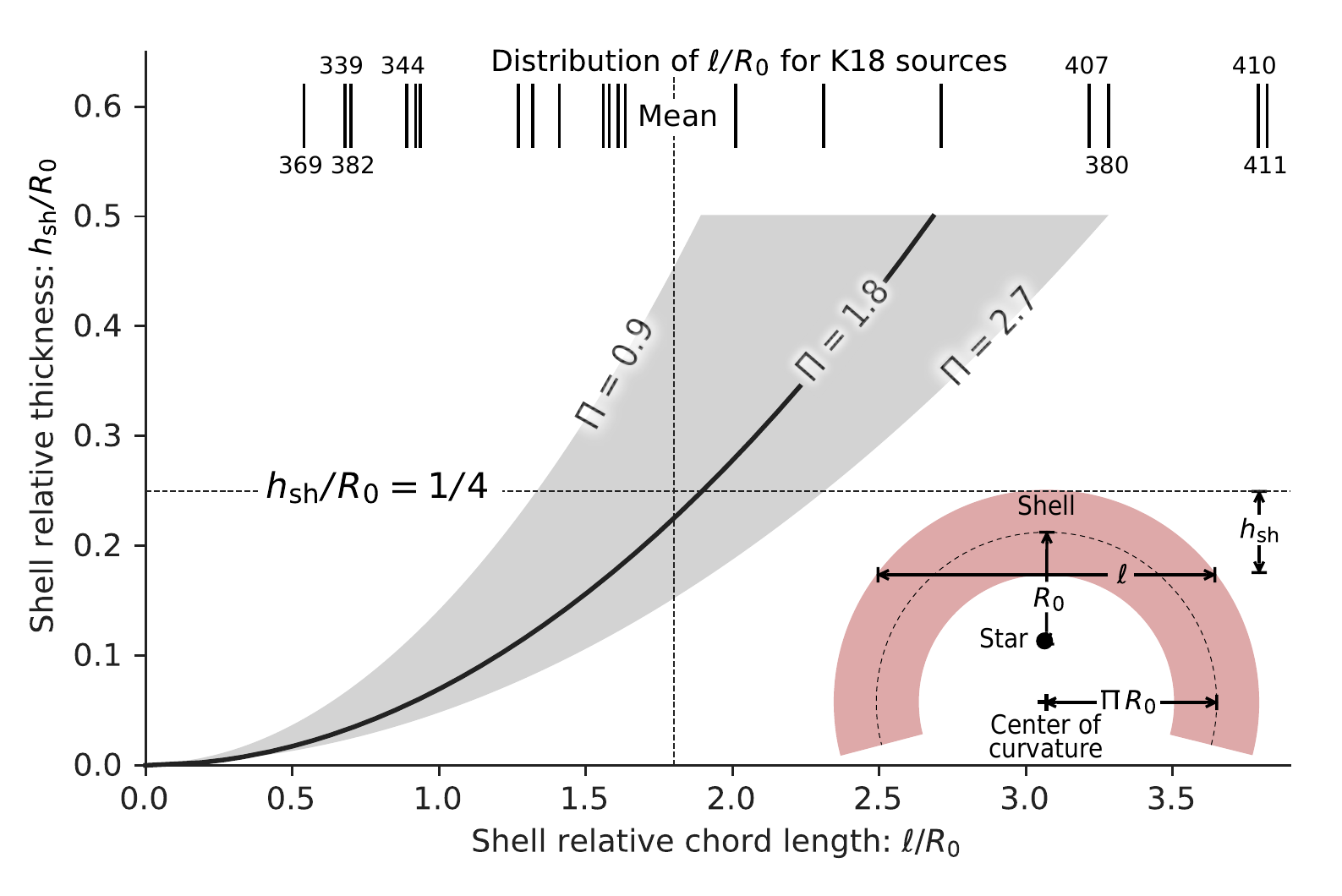}
  \caption{Relative chord length versus shell thickness.  The line and
    gray shading shows the theoretical relation (eq.~\eqref{eq:h-R0})
    for planitude of \(\Pi = 1.8 \pm 0.9\).  The short vertical lines
    at the top of the graph show the chord lengths for each of the 20
    K18 sources.  The sources with the four smallest and four largest
    values of \(\ell/R_0\) are individually labelled. The mean value
    of \(\ell / R_0 = 1.8\) is indicated by a dashed line, which
    corresponds to \(H/R_0 = 0.25\), which is the common value that we
    use in deriving \(\eta\) and therefore \(\dot{M}\) by our method.
    Inset diagram at lower right shows the geometry that leads to
    eq.~\eqref{eq:h-R0}.  }
  \label{fig:H-versus-ell}
\end{figure}

From comparing the two panels in
Figure~\ref{fig:mass-loss-vs-luminosity} it is clear that there is a
broad average agreement between the \(\dot{M}\) values from the two
methods.  Nevertheless, there is considerable disparity for individual
sources.  Obviously this is to be expected for sources where the
\(\tau\)--\(\eta\shell\) method implies radiation support, since the
K18 method assumes stellar wind support in all cases, and so has no
way of identifying these.  However, discrepancies remain even for
sources that are well above the diagonal line in
Figure~\ref{fig:All-sources-eta-tau}.  This is shown more clearly in
Figure~\ref{fig:mass-loss-comparison}, where we compare the two
mass-loss determinations. Plot symbols are colored according to the
physical size of the bow, \(R_0\), and radiation-support candidates
are shown fainter.  It is apparent that there is only weak correlation
between the two techniques (Pearson correlation coefficient
\(r = 0.67\)).  Interestingly, the smaller bows (red/orange shading)
show much better agreement than the larger bows (yellow shading).  The
five bows with \(R_0 > \SI{0.4}{pc}\) (339, 344, 369, 381, 382) show a
difference of nearly an order of magnitude, in the sense that our
method consistently predicts lower mass-loss rates than K18.  The
error ellipse for the combined observational errors (see
Tab.~\ref{tab:error-ellipse} in App.~\ref{sec:comb-uncert-covar}) is
highly elongated along the leading diagonal in
Figure~\ref{fig:mass-loss-comparison} due to the fact that most of the
observational uncertainties effect both mass-loss methods in a similar
way.  It is therefore unlikely that observational errors are a
significant contribution to the \textit{difference} in the two
\(\dot M\) methods, which must instead be due to the systematic model
uncertainties.  These are represented in the figure by the vertical
error bars, which show the effect of a factor of three uncertainty in
\(\eta\shell\) due to variations in shell thickness, dust opacity, and
shell gas temperature.  It can be seen that systematic errors of this
magnitude could indeed account for the observed differences, but the
questions remain: which parameter is causing the problem? and can we
correct for it?

Although we have seen (Fig.~\ref{fig:mass-loss-comparison}) that the
mass-loss discrepancy increases with \(R_0\), it is hard to understand
why the physical size per~se might cause this.  We have therefore
looked at the correlations between all of the observed and derived
quantities, showing some of the more interesting results in
Figure~\ref{fig:correlations}.  Each row of plots shows the dependence
(\(y\) axis) of the two different derived mass-loss rates, and their
ratio, on different parameters (\(x\) axis) of the star (leftmost four
columns) and bow shell (remaining six columns).  The most significant
correlations are marked in red with the value of the Pearson
correlation coefficient, \(r\) (calculated after taking the log of all
quantities).  The values shown with dark text have \(r^2 > 0.5\),
which means that at least 50\% of the variance in the \(y\) quantity
is ``explained'' by the variance in the \(x\) quantity (although this
cannot necessarily be taken to imply causation in either direction
since it might be due to the fact that both \(x\) and \(y\) are
partially determined by a third quantity, \(z\)). For some
correlations, we also mark the linear regression slope, \(m\).  Since
we are working in logarithmic space, this is equal to the power index
in the relation \(y \propto x^m\).  Some selected weaker correlations
are also shown (fainter text).

For our \(\eta\)--\(\tau\shell\) mass-loss method (top row) the only
significant correlation is a positive one with the shell infrared
luminosity \(L\IR\) (rightmost column), with slope
\(m = 0.6 \pm 0.1\).  For the K18 mass-loss method (middle row), the
situation is very different, showing significant correlation with a
trio of quantities: a positive correlation with bow radius, \(R_0\),
and negative correlations with relative chord length, \(\ell/R_0\),
and radiation field at the shell, \(U\).  The same three correlations
(with inverted sense) are seen for the ratio of the two methods
(bottom row).  This is due to the fact that the
\(\eta\)--\(\tau\shell\) method shows much shallower slopes in its
(weak) correlations with these quantities.  As an example, we show the
values for \(\ell/R_0\) on the figure (seventh column):
\(m = -2.3 \pm 0.4\) for the K18 method but \(m = -0.7 \pm 0.4\) for
our method, which leaves a significant slope in the ratio of
\(m = 1.6 \pm 0.4\).  For the shell luminosity, on the other hand, the
ratio does not show any significant correlation at all.  This is
because both methods have essentially the same slope in their
correlation with \(L\IR\): \(m = 0.6 \pm 0.1\) for our method, and a
weak correlation with \(m = 0.5 \pm 0.3\) for the K18 method.


Out of the three quantities, \(R_0\), \(\ell/R_0\), and \(U\) (which
are highly correlated between themselves) we will concentrate on the
relative chord length \(\ell/R_0\) as the most likely culprit for the
discrepancy between the two mass-loss estimates.  This is because it
contains observational information that is used in the K18 method
(step K\ref{K2} of \S~\ref{app:bow-shock-data}), but is \textit{not
  used} in the \(\tau\)--\(\eta\shell\) method.  The closest
equivalent of \(\ell\) in our method is the shell thickness,
\(h\shell\), which enters in step~P\ref{P2} of
\S~\ref{sec:energy-trapp-vers}, but crucially we use a fixed fraction
of the bow radius, \(h\shell / R_0 = 0.25\), rather than basing it on
any observations.  In principle, we could derive \(h\shell\) from
\(\ell\) if we knew the radius of curvature of the shell,
\(R_{\text{c}}\), or equivalently the planitude:
\(\Pi = R_{\text{c}} / R_0\) \citep{Tarango-Yong:2018a}.  The idealized
geometry is illustrated in the inset to Figure~\ref{fig:H-versus-ell},
from which we find
\begin{equation}
  \label{eq:h-R0}
  \frac{h\shell}{R_0} = \frac{1}{8\Pi} \left(\frac{\ell}{R_0}\right)^2 \ .
\end{equation}
This is plotted in Figure~\ref{fig:H-versus-ell} for planitudes of
\(\Pi = 1.8 \pm 0.9\), which are typical of OB star bows (this will be
shown in detail in the upcoming Paper~IV).  Also shown in the figure
are the individual relative chord lengths for the K18 sources (rug
plot at top).  Although the mean value of \(\ell/R_0 \approx 1.8\)
corresponds closely to the \(h\shell/R_0 = 0.25\) that we have used,
the sources with \(\ell/R_0 < 1\) are predicted to have thinner shells
with \(h\shell/R_0 < 0.1\).  The four sources with the lowest values
of \(\ell/R_0\) are labeled on the figure (369, 339, 382, 344) and
these are precisely those sources that show the largest discrepancy
between the two mass loss methods
(Fig.~\ref{fig:mass-loss-comparison}).  Using a smaller value of
\(h\shell\) would increase \(\eta\shell\) (eq.~[\ref{eq:eta-shell}])
and hence \(\dot{M}\) (eq.~[\ref{eq:mdot-tau-eta}]), thereby reducing
the discrepancy between the two methods.

We choose not make this correction to the \(\tau\)--\(\eta\shell\)
method in this paper since we want to test the methods using published
data alone and without excessive fine-tuning.  However, for future
applications an empirical measurement of the shell thicknesses would
clearly help improve the reliability of the \(\tau\)--\(\eta\shell\)
method.

It is hard to say which of the two mass-loss methods is ``better'',
except that the \(\tau\)--\(\eta\shell\) method has the advantage of
being able to identify radiation-supported bows, for which the mass
loss cannot be determined.  We recommend that both be employed if
possible as a cross-check on one another.  The
\(\tau\)--\(\eta\shell\) method has the disadvantage of requiring
shell photometry at a range of wavelengths: \SIrange{8}{70}{\um} in
order to be sure of covering the bulk of the grain emission, whereas
the K18 method needs only the \SI{70}{\um} surface brightness.  On the
other hand, we suspect that this makes the \(\tau\)--\(\eta\shell\)
method less sensitive to possible variations in grain composition and
size distribution.  This is because it depends on the average grain
opacity over a broad range of ultraviolet wavelengths, rather than the
emissivity in a single infrared band.  Additionally, many bows are
weak emitters at \SI{70}{\um}, whereas the background emission from
the PDRs surrounding \hii{} regions is very bright and variable.  The
contrast of the bow against the background is usually much higher at
\SI{24}{\um}, but K18 avoid this waveband on the grounds that it may
be dominated by emission from stochastic heating of very small grains.
We maintain that this avoidance is over-conservative (see
\S~\ref{sec:stoch-single-phot}), since (i)~the high circumstellar
radiation fields (\(U > 1000\)) mean that stochastic heating is only
relevant for ultra-small nanometer-sized grains \citep{Draine:2001a},
and (ii)~the most important class of such grains in the general ISM is
PAHs, and these are destroyed by the ionizing radiation from O~stars
\citep{Desert:1990a}.  Although PAHs may survive in bows around
B~stars, which lack the energetic photons (\(h\nu > \SI{40}{eV}\))
that most efficiently destroy them \citep{Lebouteiller:2007a}, their
contribution to the \SI{24}{\um} emissivity is likely to be small
compared with the slightly larger grains (\(> \SI{5}{nm}\)), which are
in thermal equilibrium with the radiation field.  It may therefore be
worth modifying the K18 method to use the \SI{24}{\um} surface
brightness instead of \SI{70}{\um}.  In the case of the
\(\tau\)--\(\eta\shell\) method, this is all irrelevant since
equation~\eqref{eq:tau-empirical} is valid in a time-averaged sense,
irrespective of whether stochastic heating is important or not.

\subsection{Common characteristics of radiation-supported candidates}
\label{sec:comm-prop-cand}

For five of the K18 sources (380, 407, 409, 410, 411), our method of
\S~\ref{sec:mass-loss-determ} gives only an upper limit to the wind
mass-loss rate if one assumes a factor-three uncertainty in
\(\eta\shell\), raising the possibility that they may be
radiation-supported instead of wind-supported (see Paper~I).  It is
notable that these sources are among the smallest in the K18 sample,
all with \(R_0 < \SI{0.1}{pc}\).  The two Orion Nebula bows, LP~Ori
and \thD{} also fall into this category, and they are smaller still,
with \(R_0 < \SI{0.01}{pc}\).  Six of these 7 stars also occupy a
narrow range of spectral type\footnote{%
  However, the statistical significance is not strong.  For
  instance, out of the twenty K18 sources, 7 have spectral type O8 or
  earlier.  Therefore, taking as the null hypothesis that the 5
  sources were randomly selected from the 20, then the chance that
  they should all be O9 or later is
  \(p = (1 - \frac{7}{20})^5 = 0.116\), which fails to meet the
  conventional \(2\sigma\) significance level of \(p = 0.05\),
  thereby lending only weak support for rejection of the null.} %
between O9 and B0, corresponding to
\(T_{\text{eff}} = \text{\SIrange{28}{31}{kK}}\) (LP~Ori is a much
cooler B2 star with \(T_{\text{eff}} = \SI{20}{kK}\),
\citealp{Petit:2008a, Alecian:2013a}).

The small bow sizes are fully consistent with the expectations from
the theory developed in Paper~I.  Radiation-supported bows are predicted
to occur either in high-density environments
(\(n > \SI{100}{cm^{-3}}\)) or for stars with particularly weak winds.
In either case, the bow is expected to be smaller than \SI{0.1}{pc}
for \(v_\infty \ge \SI{20}{km.s^{-1}}\), see Figures~2a,b and~8 of
Paper~I.\@

Another common property of these sources is that they all have high
ratios of chord length to stand-off radius, \(\ell/R_0\).  For
instance, from Figure~\ref{fig:H-versus-ell}, the four sources with
the highest \(\ell/R_0\) values are all among the five
radiation-supported candidates.\footnote{%
  In this case, the statistical significance is far higher, since
  \(p = (\frac{5}{20})^4 = 0.0039\) for the null hypothesis, giving
  strong support for its rejection.}  From equation~\eqref{eq:h-R0}
this means they also should have thicker than average shells.  This is
consistent with our finding from Paper~I's \S~3.3 that the shell tends
to be broad in the radiation-supported bow wave regime.  Furthermore,
if the larger than average \(H/R_0\) were to be taken account in the
calculation of \(\eta\shell\) (eq.~[\ref{eq:eta-shell}]), then it
would strengthen the case for radiation support in these objects.

\subsection{The anomalous B~star KGK2010~2 in Cygnus: strong wind or
  trapped ionization front?}
\label{sec:anomalous-b-star}

\begin{figure}
  \centering
  \includegraphics[width=\linewidth]{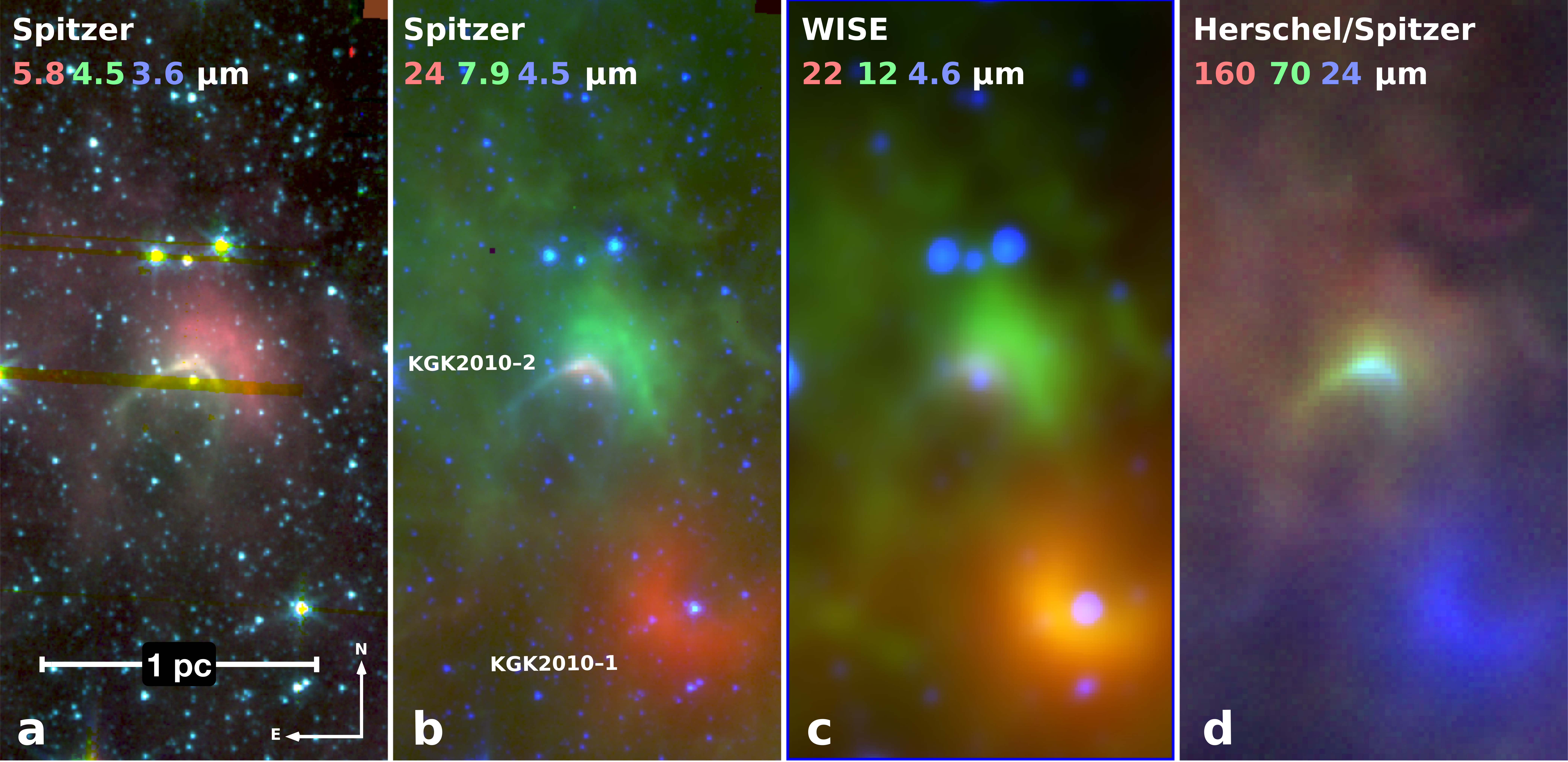}
  \caption{Two neighboring but highly contrasting bows in Cygnus:
    KGK2010~2 (source~342) and KGK2010~1 (source~341). Each panel
    shows the same \(3.6' \times 7.4'\) field, centered on equatorial
    coordinates
    \((\alpha, \delta) = 20^{\text{h}}34^{\text{m}}34.50^{\text{s}},
    \ang{+41;58;27.4}\), but in different combinations of three
    filters, as marked. }
  \label{fig:cygnus-bows}
\end{figure}

K18 source~342 is the only source that shows a significant discrepancy
\textit{above} the theoretical mass-loss rate predictions.  This is
bow shock candidate~2 in a survey of mid-infrared arcs in Cygnus~X
\citep{Kobulnicky:2010a}, and is referred to as KGK2010~2 in
subsequent works.  For its early~B spectral type,
\(\dot{M} \approx \SI{3e-9}{M_\odot.yr^{-1}}\) is expected, whereas the derived
values from both mass-loss methods (panels a and b of
Fig.~\ref{fig:mass-loss-vs-luminosity}) are about 10 times larger.
This fact was already noted in K18, who suggested that the grain
emissivity models may not be appropriate for this object.  We propose
instead to investigate the idea that the shell may be predominantly
neutral instead of ionized, which might help explain other peculiar
aspects of the source.  Figure~\ref{fig:cygnus-bows} shows infrared
images of the field containing both sources~342 and 341.  The field is
located in the Cygnus~X North region, between the DR~20 and DR~21 star
forming regions, about \SI{17}{pc} north of the center of the Cygnus
OB2 association \citep[e.g.,][]{Schneider:2016a}.  It can be seen that
the two bow shocks have very contrasting morphologies and spectra.
Source~341 has a semi-circular shape, a very diffuse outer boundary,
and is only prominent between wavelengths of \SI{12}{\um} and
\SI{24}{\um}.  These characteristics are typical of many optically
thin bows seen in fully ionized gas, such as the Ney--Allen nebula
around \thD{} \citetext{compare the bow shape in Fig.~3a of
  \citealp{Smith:2005a}}, source~406 \citetext{HD~92607, which is
  ERO~36 in Carina, \citealp{Sexton:2015b}}, and the central star of
RCW~120 \citetext{not included in current samples, but see
  \citealp{Mackey:2015a, Mackey:2016a}}.  Source~342, on the other
hand, has a more parabolic shape, a sharply defined outer boundary to
its arc, and emits strongly at all wavelengths from \SI{3.6}{\um} to
\SI{150}{\um} \citetext{see also Fig.~13 of
  \citealp{Kobulnicky:2010a}}.  The parabolic arc shape and the broad
SED are both reminiscent of LP~Ori (more details on the Orion Nebula
bows will be provided in Paper~VI), and another bow that bears some
similarities is Carina's Sickle object \citetext{ERO~21
  \citealp{Sexton:2015b}, see also \citealp{Ngoumou:2013a} and \S~4 of
  \citealp{Hartigan:2015a}}.  Note also the second outer rim in this
source, which emits primarily at \SI{5.8}{\um}, \SI{7.9}{\um}, and
\SI{12}{\um}. This appears similar to the double-bow structure seen in
IRAS~03063+5735 \citep{Kobulnicky:2012a}.  Finally, it has the coldest
dust of any of the K18 sources,\footnote{%
  Even with such a low dust temperature, adding the \SI{150}{\um} flux
  to the quadrature sum would only increase the total infrared flux by
  20\% over that given by equation~\eqref{eq:total-ir-flux}.} %
as determined from the mid-to-far infrared color temperatures
(\(T_{22/70} = \SI{70}{K}\), \(T_{70/150} = \SI{69}{K}\) from Tab.~5
of \citealp{Kobulnicky:2017a}).

If source~342 has trapped the ionization front in its shell, then
\(\eta\shell\) has been overestimated and the source should be moved
vertically down on Figure~\ref{fig:All-sources-eta-tau} in a similar
way to LP~Ori.  The exact correction factor is unknown, since it
depends on the temperature and magnetic field in the neutral shell,
but it could easily be of order 10, which would take the shell into
the radiation-supported regime.  The derived mass-loss rate would
hence be reduced from the anomalously high value of
Figure~\ref{fig:mass-loss-comparison} to essentially zero.  However,
this requires that the shell be capable of absorbing all the ionizing
photons from the central star, which implies a lower limit to the
shell optical depth: \(\tau > \tau\iftrap\), see Paper~I's
\S~3.1.\footnote{%
  Note that \(\tau\iftrap < 1\) is allowed, since \(\tau\) is the
  ultraviolet \emph{dust} optical depth of the shell, whereas it is
  the EUV \textit{gas} opacity that is most important for trapping the
  ionization front.  } %
In the radiation-supported bow wave (RBW) regime this is
\begin{equation}
  \label{eq:tau-trap-rbw}
  \text{RBW:}\quad \tau\iftrap \approx 7.1 \, \left( \frac{S_{49} \, \kappa_{600}}{L_4} \right)^{1/2} \ ,
\end{equation}
while in the wind-supported bow shock (WBS) regime it is
\begin{equation}
  \label{eq:tau-trap-wbs}
  \text{WBS:}\quad \tau\iftrap \approx 102 \, \frac{S_{49} \, \kappa_{600}}{\eta\wind\, L_4}  \ .
\end{equation}
In these equations, which are derived from equations~(8, 10, 12, 24)
of Paper~I, \(S_{49}\) is the ionizing photon luminosity of the star
in units of \SI{e49}{s^{-1}}, while \(L_4\) and \(\kappa_{600}\) are
defined in \S~\ref{sec:energy-trapp-vers} and \(\eta\wind\) in
\S~\ref{sec:mass-loss-determ}.  Within the uncertainties, the
observationally inferred spectral type of B0--B2V of source~342 is the
same as the \SI{10}{M_\odot} main-sequence star used in Papers~I
and~II, which has a ratio of ionizing to bolometric luminosity
\(S_{49}/L_4 = \num{2.1e-4}\).  Substituting into
equation~\eqref{eq:tau-trap-rbw} yields \(\tau\iftrap \approx 0.1 \)
for the RBW case, whereas the observed optical depth is more than an
order of magnitude lower: \(\tau \approx 0.005\).  Similar
difficulties arise if one assumes a wind-supported bow shock, in which
case \(\tau < \eta\wind\) by definition.  Combined with
equation~\eqref{eq:tau-trap-wbs}, this then implies a lower limit for
the wind momentum efficiency: \(\eta\wind > 0.14\).  Given that we
must always have \(\eta\shell \ge \eta\wind\)
(eq.~[\ref{eq:eta-shell-components}]), this is inconsistent with the
observed \(\eta\shell = 0.07 T_4 / \kappa_{600} h_{1/4}\), especially
since the whole point of the exercise is to reduce \(\eta\shell\) via
a reduction in \(T_4\).

Therefore, we have shown that source~342 cannot have a neutral shell
if we take the published observational data at face value.  However,
more recent spectroscopic observations revise the spectral
classification to B4~V (H.~Kobulnicky, priv.~comm.) and this would
change the picture completely.  A B4 dwarf, with effective temperature
of \SI{16700}{K} \citep[Tab.~4 of][]{Pecaut:2013a} would have a very
small ionizing luminosity. Interpolating on Table~4 of
\citet{Lanz:2007a}, we find \(S_{49} / L_4 \approx \num{4e-6}\), which
yields \(\tau\iftrap = 0.017\) from equation~\eqref{eq:tau-trap-rbw}.
At the same time, the revised spectroscopy implies a smaller \(L_*\)
by roughly a factor of 10, which increases the shell's derived
\(\tau\) by the same factor (eq.~[\ref{eq:tau-empirical}]), yielding
\(\tau \approx 0.05\).  Since we now have \(\tau > \tau\iftrap\) it is
possible that ionization front may be trapped, giving a predominately
neutral shell.  This also relies, however, on the radiation field from
the nearby Cygnus OB2 cluster being insufficient to ionize the shell.
Given that the cluster's stellar population \citep{Wright:2015a}
includes 3 WR stars and a handful of early-O supergiants, the ionizing
luminosity must be of order \SI{e51}{s^{-1}}.  In the absence of
absorption, and assuming that the true separation be no more than a
few times larger than the projected separation, the cluster stars
would contribute over 100 times greater ionizing flux at the position
of source~342 than that due to the B4 star itself.  It is therefore
necessary for the source to lie outside of the cluster's \hii{} region
in order for the shell to be neutral.  It is quite plausible that this
might be the case, since radio continuum observations \citetext{Fig.~4
  of \citealp{Wendker:1991a} and Fig.~7 of \citealp{Tung:2017a}}
indicate that the neighborhood of source~342 is a local minimum in
free-free emission between discrete \hii{} regions of the Cygnus~X
complex, while Herschel maps \citep{Schneider:2016a} show the
presence of both warm and cold dust along the line of sight and clumps
of Class~0 and~I YSOs are seen nearby \citep{Beerer:2010a}, indicative
of shielded neutral/molecular gas.

\section{Summary}
\label{sec:conclusions}

We have proposed a novel diagnostic method, the \(\tau\)-\(\eta\shell\)
diagram, for analyzing observations of stellar bow shocks around OB
stars, which allows discrimination between radiation-supported and
wind-supported bows.  Our principal results are as follows:
\begin{enumerate}[1.]
\item The UV optical depth \(\tau\) of the bow shell can be estimated
  from the observed infrared shell luminosity \(L\IR\) and stellar
  luminosity \(L_*\) as \(\tau = 2 L\IR/L_*\)
  (\S~\ref{sec:energy-trapp-vers}).
\item The shell momentum efficiency \(\eta\shell\), which is the
  fraction of stellar radiative momentum that is transferred to the
  shell either directly or indirectly (via stellar wind), can be
  estimated from \(\tau\) and the observed shell radius \(R_0\), after
  making some auxiliary assumptions about the physical conditions in
  the shell (\S~\ref{sec:energy-trapp-vers}).
\item By comparing \(\tau\) and \(\eta\shell\), it is possible to
  discriminate between three regimes (\S~\ref{sec:eta-tau-diagnostic}
  and see Paper~I): wind-supported bow shocks
  (\(\eta\shell \gg \tau\)), radiation-supported bow waves
  (\(\eta\shell \sim \tau < 1\)), and radiation-supported bow shocks
  (\(\eta\shell \sim \tau > 1\)).
\item Shells with \(\eta\shell < \num{e-3}\) are potentially in a
  fourth regime of dust waves, where gas--grain collisional coupling
  breaks down (see Paper~II).
\item By analyzing a published sample of 20 mid-infrared bow shock
  candidates \citep[K18]{Kobulnicky:2018a}, plus additional compact
  bows in Orion, we find 4 strong candidates for radiation-supported
  bow waves plus 3 marginal cases (\S\S~\ref{sec:cand-radi-supp} and
  \ref{sec:comm-prop-cand}).  No strong candidates for dust waves are
  found.
\item For wind-supported bow shocks, the stellar wind mass-loss rate
  can be found from \(\tau\) and \(\eta\shell\)
  (\S~\ref{sec:mass-loss-determ}). We compare this method with the
  previously proposed method of K18 (\S~\ref{app:bow-shock-data}) and
  suggest a correction to the dust emissivities used in the latter,
  which reduces the mass-loss rates by a factor of about 2.
\item After this correction, the two mass-loss methods agree well for
  small bows (\(R_0 \le \SI{0.3}{pc}\)), but the
  \(\tau\)--\(\eta\shell\) method under-predicts the mass-loss rate by
  up to a factor of 10 for larger bows, which also tend to have
  relatively thin shells (\S~\ref{sec:underst-diff-betw}).  This
  discrepancy could be eliminated by extending the
  \(\tau\)--\(\eta\shell\) method to include the observed shell
  thickness.
\item Disruption of the largest dust grains by rotational torques in
  the strong circumstellar radiation field is predicted to increase
  the UV opacity by about 50\% (\S~\ref{sec:evolution-grain-size}),
  although the effect could be larger if the resulting fragments are
  very small.
\item The bow source KGK2010~2 is a low-luminosity B~star with an
  anomalously high apparent mass-loss rate.  This discrepancy could be
  eliminated if the bow shell has trapped the ionization front and is
  predominantly neutral, in which case it would become yet another
  candidate for radiation support.
\end{enumerate}

\section*{Acknowledgements}
We are grateful for financial support provided by Dirección General de
Asuntos del Personal Académico, Universidad Nacional Autónoma de
México, through grant Programa de Apoyo a Proyectos de Investigación e
Inovación Tecnológica IN107019.  This work has made extensive use of
Python language libraries from the SciPy \citep{Jones:2001a}, AstroPy
\citep{Astropy-Collaboration:2013a, Astropy-Collaboration:2018a} and
Pandas \citep{McKinney:2010a} projects.  We thank Chip Kobulnicky for
sharing results in advance of publication and Thiem Hoang for useful
discussions.

\bibliographystyle{mnras}
\bibliography{bowshocks-biblio}
\appendix
\section{Covariance of derived quantities from propagation of observational uncertainties}
\label{sec:comb-uncert-covar}

\begin{table}
  \centering
  \caption[Derived]{Propagation of observational uncertainties to derived quantities}
  \label{tab:derived-parameters}
  \setlength\tabcolsep{3pt}
  \begin{tabular}{l S S S S S S}
    \toprule
    \(x_i\) & {\(\sigma_{x_i}\)}
    & {\(J_{x_i} (\tau)\)} & {\(J_{x_i} (\eta\shell)\)}
    & {\(J_{x_i} (\dot M)\)} & {\(J_{x_i} (\dot M_{\text{K18}})\)}
    & {\(J_{x_i} (L_*)\)}
    \\
    \midrule
    \(D\)     & 0.08 & 2  & 3 & 3 & 2 & 0\\
    \(L_*\)   & 0.18 & -1 & -2 & -1 & -0.5 & 1\\
    \(F\IR\)  & 0.12 & 1 & 1 & 1 & 0 & 0\\
    \(I_{70}\) & 0.12 & 0 & 0 & 0 & 1 & 0\\
    \(\theta\) & 0.11 & 0 & 1 & 1 & 2 & 0\\
    \(\ell/R\) & 0.08 & 0 & 0 & 0 & -1 & 0\\
    \(V\wind\) & 0.18 & 0 & 0 & -1 & -1 & 0\\
    \bottomrule
  \end{tabular}
\end{table}

Even though errors in the fundamental observed quantities, \(x_i\),
are assumed independent,\footnote{%
  It is of course true that \(F\IR\) and \(I_{70}\) are not completely
  independent from one another, but this does not matter since we
  consider only derived quantities that are a function of one or the
  other, not both.} %
errors in the derived quantities, \(f_k(x_1, x_2, \dots)\), will not
necessarily be so.  For the purposes of this paper:
\begin{align}
  \label{eq:observed-and-derived}
  x_i \in &  \left\{D;\ L_*;\ F\IR;\ I_{70};\ \theta;\ \ell/R;\ V\wind \right\} \\
  f_k \in & \left\{\tau;\ \eta\shell;\ \dot M;\ \dot M_{\text{K18}};\ L_* \right\}
            \ .
\end{align}
Note that \(L_*\) appears in both lists because we use it as a graph
axis in Figure~\ref{fig:mass-loss-vs-luminosity}.  For the case where
each \(f_k\) is a simple product of powers of the \(x_i\), the
propagation of errors reduces to linear algebra of log
quantities. This is exactly true for \(\tau\) and \(\eta\shell\), but
only approximately so for \(\dot M\) and
\(\dot M_{\text{K18}}\).\footnote{%
  For \(\dot M\) it is true for \(\eta\shell \gg 1.25 \tau\) and for
  \(\dot M_{\text{K18}}\) it is true in the limit that the grain
  emissivity can be expressed as a power law in \(U\).} %
We define \(J_{x_i} (f_k)\) as the elements of the Jacobian matrix of
logarithmic derivatives \(d \ln f_k / d \ln x_i \), which are given
for our quantities in Table~\ref{tab:derived-parameters}.  Then, the
elements of the variance--covariance matrix for the derived parameters
are
\begin{equation}
  \label{eq:covariance}
  C_{k,k'} = \sum_{i} J_{x_i} (f_k) \, \sigma_{x_i}^2 \, J_{x_i} (f_{k'}) \ , 
\end{equation}
where the \(\sigma_{x_i}\) are the rms dispersions in \(x_i\),
measured in dex.  In Figure~\ref{fig:python-covar} we give example
python code for calculating this matrix, using the \(\sigma_{x_i}\)
derived in \S~\ref{sec:distance}--\ref{sec:stell-wind-veloc} for the
K18 sources (second column of Tab.~\ref{tab:derived-parameters}), with
results given in Table~\ref{tab:covariance}.  It can be seen that many
of the off-diagonal elements are of similar magnitude to the diagonal
elements, which is an indication of significant correlations between
the errors in the different parameters.

\begin{table}
  \centering
  \caption[Covariance]{Variance--covariance matrix \(C_{k,k'}\) for derived quantities}
  \label{tab:covariance}
  \setlength\tabcolsep{3pt}
  \begin{tabular}{l S S S S S }
    \toprule
    & {\(\tau\)} & {\(\eta\shell\)}
    & {\(\dot M\)} & {\(\dot M_{\text{K18}}\)} & {\(L_*\)}
    \\
    \midrule
    \(\tau\)               &  0.0724 &  0.1176 &  0.0852 &  0.0418 & -0.0324 \\
    \(\eta\shell\)         &  0.1176 &  0.2137 &  0.1489 &   0.095 & -0.0648 \\
    \(\dot M\)             &  0.0852 &  0.1489 &  0.1489 &  0.1112 & -0.0324 \\
    \(\dot M_{\text{K18}}\)&  0.0418 &   0.095 &  0.1112 &  0.1353 & -0.0162 \\
    \(L_*\)                & -0.0324 & -0.0648 & -0.0324 & -0.0162 &  0.0324 \\
    \bottomrule
  \end{tabular}
\end{table}

For any particular pair of derived quantities, \(f_m\) and \(f_{n}\),
one can find the \textit{error ellipse} that characterises the
projection of observational errors onto the \(f_m\)--\(f_{n}\) plane.
The ellipse is characterized by standard deviations along major and
minor axes, \(\sigma_a\), \(\sigma_b\), together with the angle
\(\theta_a\) between the \(f_m\) axis and the ellipse major axis.
These are given via the eigenvalues and eigenvectors of the relevant
\(2 \times 2\) submatrix of the covariance matrix:
\begin{align}
  \label{eq:error-ellipse}
  \sigma_a^2 = & \frac12 \left\{   C_{m,m} + C_{n,n}
                 + \left[ \left( C_{m,m} + C_{n,n} \right)^2
                 - 4 C_{m,n}^2 \right]^{1/2}\right\} \\
  \sigma_b^2 = & \frac12 \left\{   C_{m,m} + C_{n,n}
                 - \left[ \left( C_{m,m} + C_{n,n} \right)^2
                 - 4 C_{m,n}^2 \right]^{1/2}\right\} \\
  \theta_a = & \frac12 \arctan \left( \frac{2 C_{m,n}}{C_{m,m} - C_{n,n}} \right) \ .
\end{align}
For instance, Table~\ref{tab:error-ellipse} shows the resultant error
ellipse parameters (shown in blue on the respective graphs) for the
relations plotted in Figures~\ref{fig:All-sources-eta-tau},
\ref{fig:mass-loss-vs-luminosity}ab, and
\ref{fig:mass-loss-comparison}.

\begin{table}
  \centering
  \caption[Error ellipse]{Error ellipse parameters for particular pairs of derived quantities}
  \label{tab:error-ellipse}
  \begin{tabular}{l l S S S l}
    \toprule
    \(f_m\) & \(f_n\) &  {\(\sigma_a\)} & {\(\sigma_b\)}
    & {\(\theta_a\), \si{\degree}} & Figure   \\
    \midrule
    \(\tau\) & \(\eta\) & 0.529 & 0.077 & 60.5 & \ref{fig:All-sources-eta-tau} \\
    \(L_*\) & \(\dot M\) & 0.397 & 0.155 & -75.5 & \ref{fig:mass-loss-vs-luminosity}a \\
    \(L_*\) & \(\dot M_{\text{K18}}\) & 0.371 & 0.173 & -81.3 & \ref{fig:mass-loss-vs-luminosity}b \\
    \(\dot M_{\text{K18}}\) & \(\dot M\) & 0.503 & 0.175 & 46.7 & \ref{fig:mass-loss-comparison} \\
    \bottomrule
  \end{tabular}
\end{table}

\begin{figure}
  \centering
  \footnotesize
  \begin{verbatim}
import numpy as np
sig = np.diag([0.08, 0.18, 0.12, 0.12, 0.11, 0.08, 0.18])
J = np.array([
    [2, -1, 1, 0, 0, 0, 0],
    [3, -2, 1, 0, 1, 0, 0],
    [3, -1, 1, 0, 1, 0, -1],
    [2, -0.5, 0, 1, 2, -1, -1],
    [0, 1, 0, 0, 0, 0, 0]
])
C = J @ sig**2 @ J.T
  \end{verbatim}
  \vspace*{-\baselineskip}
  \caption{Snippet of Python code that calculates the
    covariance matrix of Table~\ref{tab:covariance}.  The last line
    implements equation~\eqref{eq:covariance} by a triple matrix
    product of the Jacobian matrix \texttt{J}, the square of a
    diagonal matrix of observational standard deviations \texttt{sig},
    and the transpose of \texttt{J}.}
  \label{fig:python-covar}
\end{figure}



\section{The Herbig Be star V750~Mon}
\label{sec:notes-part-sourc}
\label{sec:hd-53367-v750}
  
K18 source~380 (HD~53367, V750~Mon) is a Herbig Be star with spectral
type B0V--B0III and mass 12 to \SI{15}{M_\odot}, which shows
long-scale irregular photometric variability \citep{Tjin-A-Djie:2001a,
  Pogodin:2006a} together with cyclic radial velocity variations,
which are interpreted as an eccentric binary with a \SI{5}{M_\odot}
pre-main-sequence companion.  It is located outside the solar circle
and, although it was originally classified as part of the CMa OB
association at about \SI{1}{kpc} \citep{Tjin-A-Djie:2001a}, more
recent estimates put it much closer.  K18 assume a distance of
\SI{260}{pc}, based on Hipparcos measurements
\citep{van-Leeuwen:2007a}, but its Gaia DR2 parallax
\citep{Gaia-Collaboration:2016a, Gaia-Collaboration:2018a, Luri:2018a}
puts it closer still at \SI{140 \pm 30}{pc}.  This is the median and
90\% confidence interval, estimated from Bayesian
inference\footnote{\url{https://github.com/agabrown/astrometry-inference-tutorials/}.}
using an exponential density distribution with scale length of
\SI{1350}{pc} as a prior.

\citet{Quireza:2006b} report a kinematic distance to the associated
\hii{} region IC~2177 (G\num{223.70}\num{-1.90}) of \SI{1.6}{kpc},
based an LSR radio recombination line velocity of \SI{+16}{km.s^{-1}}
\citep{Quireza:2006a} and the outer Galaxy rotation curve of
\citet{Brand:1993a}.  However, given the dispersion in peculiar
velocities of star-forming clouds (\SIrange{7}{9}{km.s^{-1}},
\citealp{Stark:1984a}) and likely streaming motions of the ionized gas
(\(\approx \SI{10}{km.s^{-1}}\), \citealp{Matzner:2002a, Lee:2012a}),
this is still consistent with a much smaller distance, which would
also help in bringing the radio continuum-derived nebular electron
density into agreement with the value (\(\approx \SI{100}{cm^{-3}}\))
derived from the optical [\ion{S}{ii}] line ratio
\citep{Hawley:1978a}.

\citet{Fairlamb:2015a} derive a distance of \SI{340 \pm 60}{pc} from
combining a spectroscopically determined effective temperature,
gravity, and reddening with photometry and pre-main sequence
evolutionary tracks.  They also determine a luminosity of \SI{13000
  \pm 1000}{L_\odot}, which is only half that assumed by K18, and
would have to be even lower to bring the photometry into concordance
with the Gaia distance.  Alternatively, the luminosity could remain
the same if the total-to-selective extinction ratio were higher than
the \(R_V = 3.1\) assumed by \citet{Fairlamb:2015a}.  Taking
\(R_V = 5.5\) instead, as in Orion, would give \(A_V = 3.34\) and a
predicted \(V\) magnitude of 6.7 if we assume \(L_4 = 1.3\),
\(D = \SI{170}{pc}\) (furthest distance in Gaia 90\% confidence
interval) and \(T_{\text{eff}} = \SI{29.5}{kK}\)
\citep{Fairlamb:2015a} for a bolometric correction of \(-2.86\)
\citep{Nieva:2013a}.  The observed brightness varies between
\(V = 6.9\) and 7.2, which is still at least 20\% fainter than
predicted, but this is the best that can be achieved without rejecting
the Gaia parallax distance entirely.

A final sanity check can be performed by considering the free-free
radio continuum flux of V750~Mon's surrounding \hii{} region, which,
after converting to a luminosity, should be proportional to the total
recombination rate in the nebula and therefore, assuming
photoionization equilibrium and negligible dust absorption in the
ionized gas, also proportional to the ionizing photon luminosity of
the star.  \citet{Quireza:2006b} report a flux density of \SI{6}{Jy}
at \SI{8.6}{Ghz} for G\num{223.70}\num{-1.90}, as compared with a flux
density of \SI{260}{Jy} for the Orion Nebula using the same
instrumental setup.  Assuming an ionizing photon luminosity of
\(S_{49} = 1\) and distance \SI{410}{pc} for the ionizing Trapezium
stars in Orion, therefore implies
\(S_{49} = 0.0027 (D/\SI{140}{pc})^2\) for V750~Mon.  Using an
ionizing flux of \SI{3.2e22}{cm^{-2}.s^{-1}} from the curves in Fig.~4
of \citet{Sternberg:2003a}, this translates to a bolometric luminosity
of \(L_4 = 0.96 (D/\SI{140}{pc})^2\), which is consistent with the
\citet{Fairlamb:2015a} value if we take a distance towards the high
end of the Gaia range.







\bsp	
\label{lastpage}
\end{document}
